\documentclass[preprint,preprintnumbers,amsmath,amssymb]{revtex4}
\usepackage{graphicx}
\usepackage{bm}

\usepackage{mhchem}
\usepackage{bm}
\usepackage{amsmath}
\usepackage{amsfonts}
\usepackage{amssymb}

\usepackage{breqn}

\usepackage{balance}

\usepackage{times,mathptmx}

\newcommand{\be}{\begin{equation}}
\newcommand{\ee}{\end{equation}}
\newcommand{\bea}{\begin{eqnarray}}
\newcommand{\eea}{\end{eqnarray}}

\begin{document}
\title{Cooperativity flows and Shear-Bandings: a statistical field theory approach}
\author{R. Benzi$^{1}$, M. Sbragaglia$^{1}$, M. Bernaschi$^{2}$, S. Succi$^{2}$ and F. Toschi$^{2,3}$ \\
$^{1}$ Department of Physics and  INFN, University of ``Tor Vergata'', Via della Ricerca Scientifica 1, 00133 Rome, Italy\\
$^{2}$ Istituto per le Applicazioni del Calcolo CNR, Via dei Taurini 19, 00185 Rome, Italy\\
$^{3}$ Department of Physics and Department of Mathematics and Computer Science and J.M. Burgerscentrum, Eindhoven University of Technology, 5600 MB, Eindhoven, Netherlands}

\begin{abstract} 
Cooperativity effects have been proposed to explain the non-local rheology in the dynamics of soft jammed systems. Based on the analysis of the free-energy model proposed by L. Bocquet, A. Colin \& A. Ajdari ({\em Phys. Rev. Lett.} {\bf 103}, 036001 (2009)), we show that cooperativity effects resulting from the non-local nature of the fluidity (inverse viscosity), are intimately related to the emergence of shear-banding configurations. This connection materializes through the onset of inhomogeneous compact solutions (compactons), wherein the fluidity is confined to finite-support subregions of the flow and strictly zero elsewhere. Compactons coexistence with regions of zero fluidity (``non-flowing vacuum'') is shown to be stabilized by the presence of mechanical noise, which ultimately shapes up the equilibrium distribution of the fluidity field, the latter acting as an order parameter for the flow-noflow transitions occurring in the material.
\end{abstract}

\maketitle
\section{Introduction}
Soft amorphous materials, including emulsions, foams, microgels, and others, display complex flow properties, intermediate between the solid and the liquid state of matter~\cite{Larson,Voigtmann,Coussot}.  The response of such systems to an external shear stress shows that  beyond a critical yield stress, $\sigma_Y$, they flow like liquids whereas, below $\sigma_Y$, they remain jammed and respond elastically. These materials are commonly denoted as yield-stress fluids (YSF). The yielding behavior makes such systems both interesting for applications and challenging from the fundamental point of view of out-of-equilibrium statistical mechanics~\cite{Larson,Voigtmann,Coussot}. Experimental studies based on the steady state rheology, and the behaviour in the proximity of the yield stress, have recently pointed out the existence of two apparently distinct classes of YSF's. The first class~\cite{Ragouilliaux07,Manneville2011a,Ovarlez10}, usually associated with underlying repulsive interactions~\cite{Becu}, is characterized by a {\it continuous} and smooth transition between solid-like and liquid-like behaviour, as the stress is increased above $\sigma_Y$. The corresponding rheological response is usually well fitted by generalised Herschel-Bulkley (HB) flow-curves, corresponding to a monotonous trend of the mean value of the stress $\sigma$ as a function of the imposed shear $S$:
\begin{equation}\label{eq:HB}
\sigma = \sigma_Y + A S^n.
\end{equation}
In the above equation, $A$ is a plastic viscosity and $n$ is a non-universal scaling exponent generally below 1, and often close to 0.5~\cite{Princen89,Marze08,Denkov09}. Under the condition of a constant imposed shear, the steady flowing state of these materials, often denoted as ``simple'' YSF, is homogeneous in space, even at vanishingly small imposed shear. The second class of YSF's displays a {\it discontinuous} transition from solid-like to liquid-like behaviour upon increasing an imposed stress above the critical threshold~\cite{Coussot02b,Becu,Paredes,Ovarlez08,Moller08}: the shear jumps discontinuously from a zero value below the threshold stress to a finite value just above it. This may also be regarded  as a discontinuous jump in the viscosity, from formally infinity, just below threshold, to a finite value just above it, a phenomenon often indicated by using the term {\it viscosity bifurcation}~\cite{Coussot02b,Ragouilliaux07,BonnCoussot,Coussot}. This scenario underlies the formation of shear-bands, where fluidized zones can coexist with solid-like regions, under the conditions of a constant imposed shear. These kind of materials are usually referred to as "viscosity-bifurcating" YSF's. Within the above two classes of materials, only viscosity-bifurcating  YSF's appear to support shear-banding as a permanent response to a steady imposed shear.\\
Developing predictive theories for the deformation and flow of amorphous materials, study the formation of shear-bands, as well as identifying the correct structural variables for these systems remains an ongoing challenge: in the absence of a comprehensive microscopic theory, various mesoscopic models have been vigorously pursued in the literature~\cite{Sollich1,Sollich2,Sollich3,Fielding1,Fielding2,Langer,Pouliquen,Derec,Picard01,Picard,Mansard11,Nicolas13,Mansard13}. It is also commonly accepted that as soon as the flow becomes heterogeneous, a description of the rheological behaviour solely in terms of the flow-curve is insufficient~\cite{Goyon08,Bocquet09,Goyon10,Geraud13,Kamrin12,Amon12,Katgert10,Sbragaglia12,Mansard13}. This has been illustrated in the experimental work of Goyon {\it et al.}~\cite{Goyon08}, who showed that a single flow-curve is not able to account for the flow profile of a concentrated emulsion in a microfluidic channel. That finding triggered the need to properly bridge between {\it local} and global rheology of the soft-glassy. Specifically, Goyon {\it et al.}~\cite{Goyon08} introduced the concept of {\it spatial cooperativity} length $\xi_c$, by postulating that the fluidity, $f = S/\sigma$, defined as the ratio between shear $S$ and shear stress $\sigma$, follows a non-local diffusion-relaxation equation when it deviates from its bulk value
\begin{equation}
\label{eq:fluidity}
\xi_c^2 \Delta f(\vec{r}) + f_b(\sigma(\vec{r}))-f(\vec{r}) =0.
\end{equation}
The quantity $f_b=f_b(\sigma(\vec{r}))$ is the bulk fluidity, {\em i.e.,} the value of the fluidity in the absence of spatial heterogeneities. The bulk fluidity depends upon the local stress $\sigma$, whereas $f=f(\vec{r})$ depends upon the position in space. Its value is equal to $f_b$ without the effect of cooperativity ($\xi_c=0$). The spatial cooperativity $\xi_c$ has been shown to be in the order of few times the size of the elementary microstructural constituents~\cite{Goyon08,Goyon10,Geraud13,Sbragaglia12,JFM}. The non-local equation \eqref{eq:fluidity} has been justified~\cite{Bocquet09} by using a kinetic model for the elastoplastic (KEP) dynamics of a jammed material, which takes the form of a nonlocal kinetic equation for the stress distribution function. Such model predicts nonlocal equations of the form \eqref{eq:fluidity}, plus an equation predicting a proportionality between the fluidity and the rate of plastic events $\Gamma$,
\begin{equation}
\label{eq:proportionality}
f = \frac{S}{\sigma} \propto \Gamma.
\end{equation}
An interesting interpretation of the diffusion equation \eqref{eq:fluidity} has been put forward in~\cite{Bocquet09}, based on a ``free-energy'' for the rate of plastic events $\Gamma$
\begin{equation}\label{eq:free}
F(\Gamma)= \int \left(\omega(\Gamma,\sigma)+\frac{\xi^2}{2}|{\bm \nabla} \Gamma|^2 \right) d^3x
\end{equation}
where the bulk potential $\omega(\Gamma,\sigma)$ embeds the information about the non-linear rheology, whereas an {\it inhomogeneity  parameter} $\xi$ appears as a multiplicative factor in front of the gradient terms and may be directly related to the cooperativity length $\xi_c$. Upon minimization, the free-energy formulation \eqref{eq:free} leads to the HB form of a flow-curve \eqref{eq:HB} in case of homogeneous flow, and to the fluidity equation \eqref{eq:fluidity} in case of non-homogeneous solution. As pointed out in~\cite{Bocquet09}, the square-gradient expression of the free-energy \eqref{eq:free} opens up a far-reaching connection to shear-banding. In this picture, shear-banding would correspond to a first order phase transition scenario: i.e., the spatial coexistence of two states of different fluidity values for the same shear stress. It is the purpose of the present paper to elaborate on these concepts, proposing a new viewpoint on the formation of shear-bands in soft matter systems. Shear-banding will be linked to the onset of compact configurations of the fluidity field (compactons), which correspond to the local minima of the free-energy functional. More specifically, we start from the non-local formulation due to Bocquet {\it et al.}~\cite{Bocquet09}, whose main idea is to introduce the fluidity as the order parameter of a corresponding free-energy. Free-energy minimization leads to a non-linear Helmoltz equation, whose solutions describe spatial relaxation to a uniform background fluidity, corresponding to homogeneous bulk rheology, typically in HB form. We also inspect the dynamics of the system subject to stochastic perturbations, arguably related to a form of mechanical noise in the system.  None of these concepts is brand new in the literature, although the deep and non trivial consequences shown in this paper definitely are. We are going to show that the cooperative effects, resulting from the non-local nature of the fluidity, are intimately related to the emergence of shear-banding configurations. This connection materializes through the onset of inhomogeneous compact solutions, wherein the fluidity is confined to finite-support sub-regions of the flow and zero elsewhere (non-flowing vacuum). In the absence of noise, compactons attain lower free-energy minima than the homogeneous HB solution, thereby realizing metastable shear-bands. Indeed, since the non-flowing region is unstable, such shear-bands cannot survive indefinitely: depending on the initial conditions, shear-bands may or may not occur, but even when they do, in the time-asymptotic limit they surrender to homogeneous HB configurations. The picture takes a drastic upturn once noise is taken into account. Here, a new qualitative effect arises: the effective free-energy, including renormalized fluctuations, develops two local minima, corresponding to a stable coexistence of compactons and non-flowing vacua. Under such conditions, permanent shear-banding solutions can indeed be observed. The emerging picture is conceptually sound and appealing: compactons represent natural carriers of shear-bands. They attain local minimization of the free-energy functional, but in the absence of noise they ultimately surrender to homogeneous configurations due to the instability of non-flowing vacuum.\\
We remark that the "minimization of the free energy" is a (quite) strong assumption that can be hardly justified at this stage. Nevertheless, we shall show that this idea has far reaching consequences. In particular, as a consequence of the noise, there is a renormalization of the free energy leading to an unstable rheological branch. Interestingly, the renormalization effects can be computed analytically within the framework of a Hartree-like approximation (see Appendix \ref{appendix2}). The analysis can be performed for a very simple structure of the noise, namely a white noise in space and time. We may argue that such a choice of the noise can be considered quite unrealistic and in principle the noise could depend on a external forcing or the rate of energy dissipation and it may be correlated in space and/or in time. Our choice has been done to maintain the whole theory analytically manageable without introducing a larger set of independent parameters. Based on our results, future investigations may include more realistic structures of the noise term and the relative renormalization effects.\\
The the paper is organized as follows: in Section \ref{sec:modeleq} we describe the basic features of the free-energy model, specializing to the geometry of the Couette flow; in Section \ref{sec:model} we explore the energy landscape of the model and we describe the compact solutions; in Section \ref{sec:bands} we study the free-energy of both the HB and compact solutions and the resulting velocity profiles; the stability of HB and compact solutions, as well as the aging properties of the model, are the subject of Section \ref{sec:aging}. In Section \ref{sec:noise} we analyze the behaviour of the system under stochastic perturbations (noise). The far-reaching implications of the stochastic perturbations for the geometry of the Couette flow are illustrated in Section \ref{sec:couette}. In Section \ref{sec:discussion}, we offer a discussion with comparisons between our model and other earlier works that report about the formation of permanent shear-bands. Conclusions follow in Section \ref{sec:conclusions}. Technical details for the stability of compact solutions and the self-consistent field approximation in presence of noise are reported in Appendices \ref{appendix1}-\ref{appendix2}.

\section{Model Equations}\label{sec:modeleq}

We are interested in the dynamics of a soft-glassy system within two walls at $y=0$ and $y=L$ driven with a velocity difference $\Delta U=S L$ at the boundaries, with $S$ the imposed shear. Using a HB relation with $n=1/2$, we rescale the original variables according to the following transformations:
\begin{equation}\label{scaling}
y \rightarrow \frac{y}{L} \hspace{.2in} t \rightarrow \left(\frac{\sigma_Y}{A} \right)^2 t \hspace{.2in} \sigma \rightarrow \frac{\sigma}{\sigma_Y}.
\end{equation}
As a consequence, the HB relation \eqref{eq:HB} retains a unitary yield stress
\begin{equation}
\sigma=1+S^{1/2}.
\end{equation}
Following Bocquet {\it et al.}~\cite{Bocquet09}, we write equation \eqref{eq:free} as
\begin{equation}\label{Ffm}
\begin{split}
F[f] = & \int_0^1 \left(\omega(\Gamma,\sigma)+\frac{\xi^2}{2}|{\bm \nabla} \Gamma|^2 \right)_{\Gamma=f} dy =\\
& \int_0^1 \left[-\frac{1}{2} m(\sigma) f^2 + \frac{2}{5} f^{5/2} + \frac{1}{2} \xi^2 (\partial_y f)^2 \right]
dy
\end{split}
\end{equation}
where
\begin{equation}\label{eq:msigma}
m(\sigma) \equiv \frac{(\sigma-1)}{\sigma^{1/2}}.
\end{equation}
Equations \eqref{Ffm}-\eqref{eq:msigma} are the starting point of our investigations. In case of a spatially homogeneous solution, the minimum of $F[f]$ is given by
\begin{equation}
\label{eq:bulk}
f=m^2=\frac{(\sigma-1)^2}{\sigma}
\end{equation}
that is the HB solution. For spatially non-homogeneous solutions, upon linearizing around $f_b \equiv m^2$, we get the equation
\begin{equation}\label{eq:linear}
\xi_r^2 \partial_{yy} f - f + f_b = 0
\end{equation}
which directly maps into \eqref{eq:fluidity} with a squared cooperativity length $\xi^2_c = 2 \xi^2/m$ diverging at the yield stress (as $\sigma \rightarrow 1$, $m \rightarrow 0$)~\cite{Bocquet09}. Let us remark that the value of $m=(\sigma-1)/\sigma^{1/2}$ in the above equations is consistent with the definition of the fluidity $f=S/\sigma$. In a Couette flow, assuming as usual that the stress is constant in space, we can obtain a constraint between the imposed shear and the space-averaged fluidity
\begin{equation}\label{vincolobeginning}
S= \sigma \int_0^1 f(y)\, dy.
\end{equation}

\section{Energy landscape at imposed stress: Herschel-Bulkley {\it vs.} compact solutions}\label{sec:model}

\begin{figure}[t!]
\begin{center}
\includegraphics[width=0.6\textwidth]{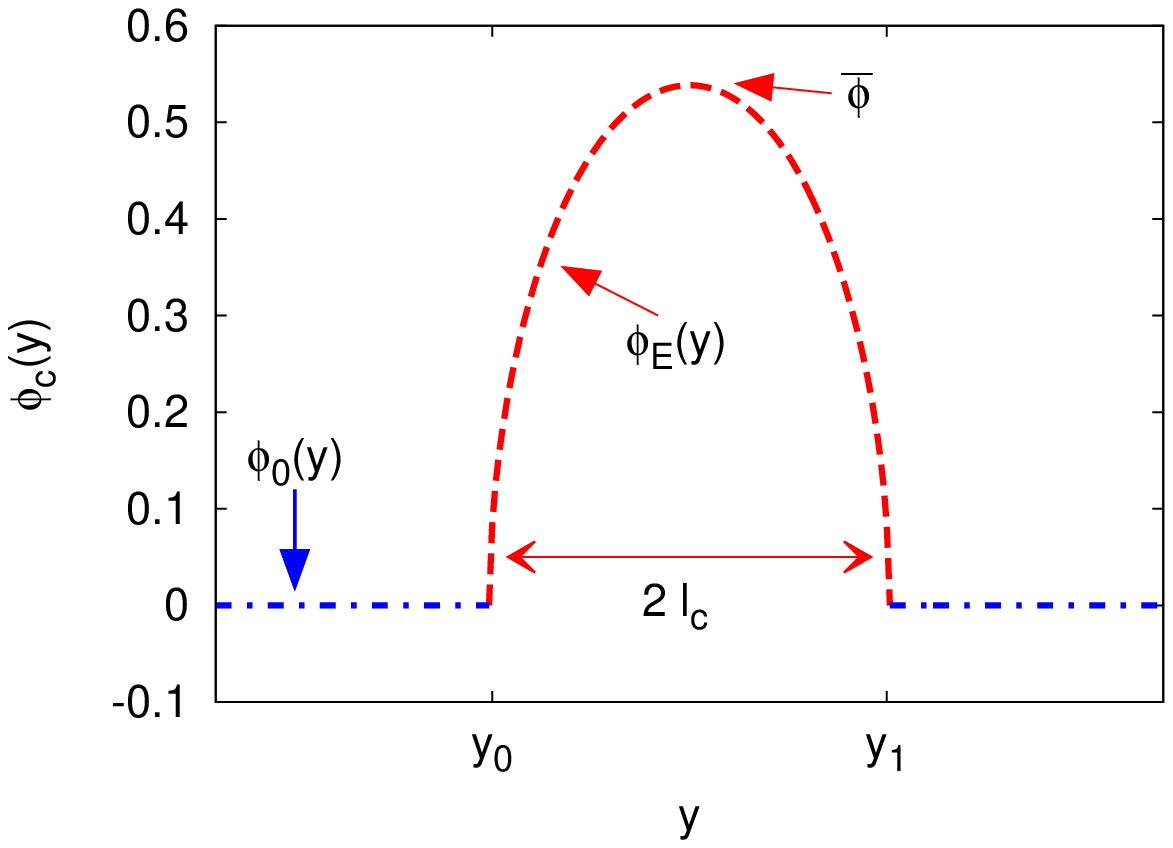}
\includegraphics[width=0.6\textwidth]{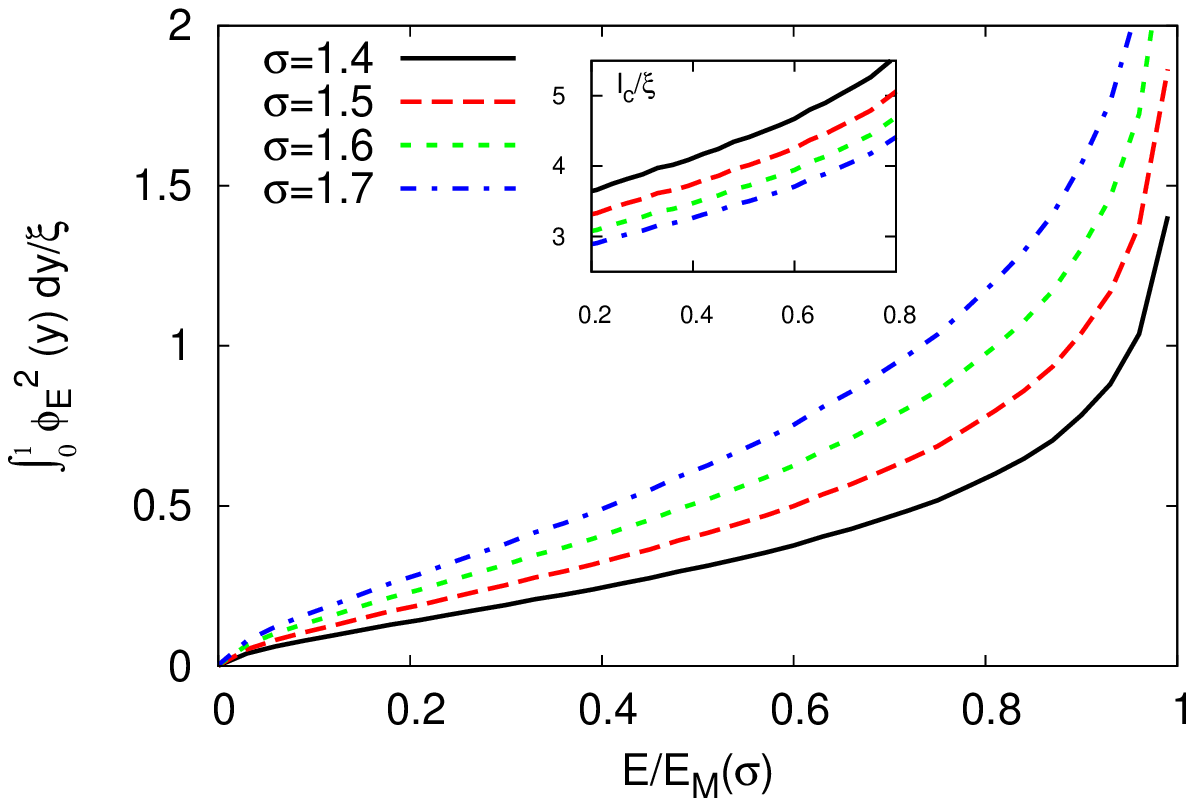}
\end{center}
\caption{Properties of compact solutions $\phi_c$ \eqref{eq:compacton}. Top panel: a sketch of a compact solution obtained from equation \eqref{phi0}: the order parameter $\phi$ is non zero only in some compact sub-domain $[y_0,y_1]$ where the profile is identified with the compacton $\phi_E$ (red dashed). The compacton $\phi_E$ has a maximum $\phi(\bar{y})=\bar{\phi}$ in $\bar{y}=(y_0+y_1)/2$. Outside the compact sub-domain the order parameter is identically zero (blue dashed-dotted). Bottom panel: the average fluidity associated with the compacton $\phi_E$, {\em i.e.,} $\int_0^1 \phi^2_E(y)\,dy$, for different values of the imposed stress $\sigma$ in the free-energy \eqref{F}. The average fluidity is reported as a function of the constant of integration $E$ (see equation \eqref{E1}) normalized to its maximum allowed value $E_M(\sigma)=\frac{1}{20}\frac{(\sigma-1)^5}{\sigma^{5/2}} $ (see equation \eqref{EMAX} and text for details). Notice that the average fluidity is made dimensionless with respect to the inhomogeneity parameter $\xi$ of the free-energy \eqref{F}. In the inset we report the half width $l_c$ of the compacton calculated according to equation \eqref{lc} as a function of $E/E_M$, again for different values of the imposed stress $\sigma$ (same as the main panel). \label{fig:compacton}}
\end{figure}

Let us start with the analysis of the energy landscape in the system at an imposed constant stress $\sigma$. We want to specialize our analysis to those situations where the ``order parameter'' $f$ is positive definite. This request can be ensured if we set $f=\phi^2$ and use the corresponding free-energy functional, directly obtained from \eqref{Ffm} with the substitution $f=\phi^2$:
\begin{equation}\label{F}
\begin{split}
F[\phi] = & \int_0^1 {L}(\phi, \partial_y \phi) \, d y=\\
& 2 \int_0^1 \left[-\frac{1}{4} m(\sigma) \phi^4 + \frac{1}{5} \phi^4 |\phi|+ \xi^2\phi^2 (\partial_y \phi)^2 \right] d y.
\end{split}
\end{equation}
Note that the term $f^{5/2}$ has been rewritten as $\phi^4 |\phi|$ to guarantee $F[\phi]>0$ in the limit $\phi \rightarrow \pm \infty$. We then look for the local extrema of the free-energy functional \eqref{F}. The variational equation $\frac{\delta F}{\delta \phi} = 0$
gives:
\begin{equation}\label{phi0}
2 \xi^2 \phi^2 \partial_{yy} \phi + 2 \xi^2 \phi (\partial_y \phi)^2 + m(\sigma)
\phi^3 - \phi^3 |\phi| = 0.
\end{equation}
Equation \eqref{phi0} exhibits solutions with constant order parameter ($\phi_m=m$), {\em i.e.,} the bulk HB solutions. However, also {\it compact} solutions are possible
\begin{equation}\label{eq:compacton}
\phi_c(y) = \begin{cases} \begin{split}  & \phi_{0}(y)  & y \le y_0; y \ge y_1\\
                           & \phi_E(y)  & y_0 \le y \le y_1 \end{split} \end{cases}
\end{equation}
with non zero values of the order parameter only in some compact sub-domain, say $[y_0,y_1]$, and zero elsewhere (see the top panel of Figure \ref{fig:compacton} for a sketch). In the above, $\phi_{0}(y)=0$ is a ``vacuum'' field, corresponding to a zero-fluidity (non-flowing) state. The structure of the compact solution in the region where the order parameter $\phi$ is different from zero, {\em i.e.,} the function $\phi_E(y)$ (hereafter named {\it compacton}), can be further characterized by taking one quadrature of equation \eqref{phi0}
\begin{equation}\label{E1}
\xi^2 \phi^2 (\partial_y \phi)^2 = E - \frac{m(\sigma)}{4} \phi^4 +
\frac{1}{5} \phi^5
\end{equation}
where $E$ is a positive constant and where, for the sake of simplicity, we assume $\phi>0$. By increasing $\phi$ from zero to larger values, and for sufficiently small values of $E$ \footnote{The rhs of \eqref{E1} has a minimum in $\phi=m$. The condition on non-negativity of the rhs of \eqref{E1} in $\phi=m$ leads to the condition $E < m^5/20$, hence equation \eqref{EMAX} because $m(\sigma)=(\sigma-1)/\sigma^{1/2}$.}
\begin{equation}\label{EMAX}
E \le E_{M}(\sigma) = \frac{1}{20}\frac{(\sigma-1)^5}{\sigma^{5/2}}
\end{equation}
the r.h.s. of \eqref{E1} first becomes zero for a value of the order parameter denoted with $\phi = \bar{\phi} < m$. Then, we can look for a solution of \eqref{E1} localized in the interval $[y_0,y_1]$, with $\phi(\bar{y})=\bar{\phi}$ and $\bar{y}=(y_0+y_1)/2$. Based on the structure of the exact solution
\begin{equation}\label{exact}
\int_{0}^{\phi} \frac{\psi}{\sqrt{ E - \frac{m(\sigma)}{4} \psi^4 + \frac{1}{5} \psi^5}} \, d \psi = \frac{y-y_0}{\xi}
\end{equation}
we obtain the characteristic size of the region, $|y_1-y_0| = 2l_c$,
where the order parameter is different from zero
\begin{equation}\label{lc}
l_c = \xi \int_{0}^{\bar{\phi}} d\psi \frac{\psi}{\sqrt{ E - \frac{m(\sigma)}{4} \psi^4 + \frac{1}{5} \psi^5}}.
\end{equation}
It is easy to check that for $y$ close to $y_0$, {\em i.e.,} for small
$\phi$, we have
\begin{equation}\label{local}
\phi \sim \left( \frac{E}{\xi^2} \right)^{1/4} (y-y_0)^{1/2}
\end{equation}
implying that $\partial_y \phi$ diverges as $1/(y-y_0)^{1/2}$ at small $\phi$. Importantly, equation \eqref{phi0} is well defined even at the singular points since all the divergences cancel out. It is also important to highlight that the characteristic size of the compact region \eqref{lc} scales proportionally to $\xi$, hence no compact solution can be achieved without cooperativity. Since compact solutions do not overlap, superpositions of compactons still correspond to local extrema of the free-energy. This implies that the energy in equation \eqref{E1} takes the form of a piece-wise constant function, which attains distinct non-zero values in different compactons and is zero elsewhere.


\section{Flow Profiles at imposed Shear: Linking Compactons to Shear-Bands}\label{sec:bands}

\begin{figure}[t!]
\begin{center}
\includegraphics[width=0.6\textwidth]{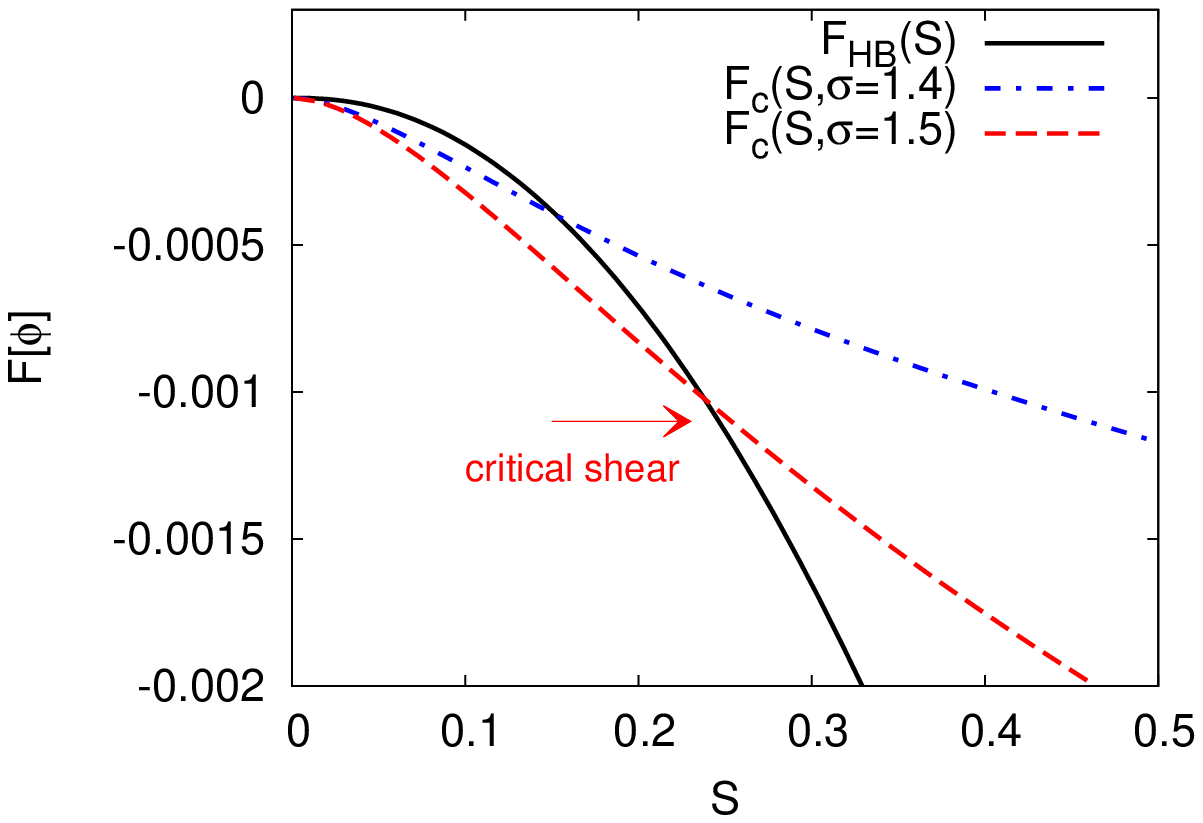}
\includegraphics[width=0.6\textwidth]{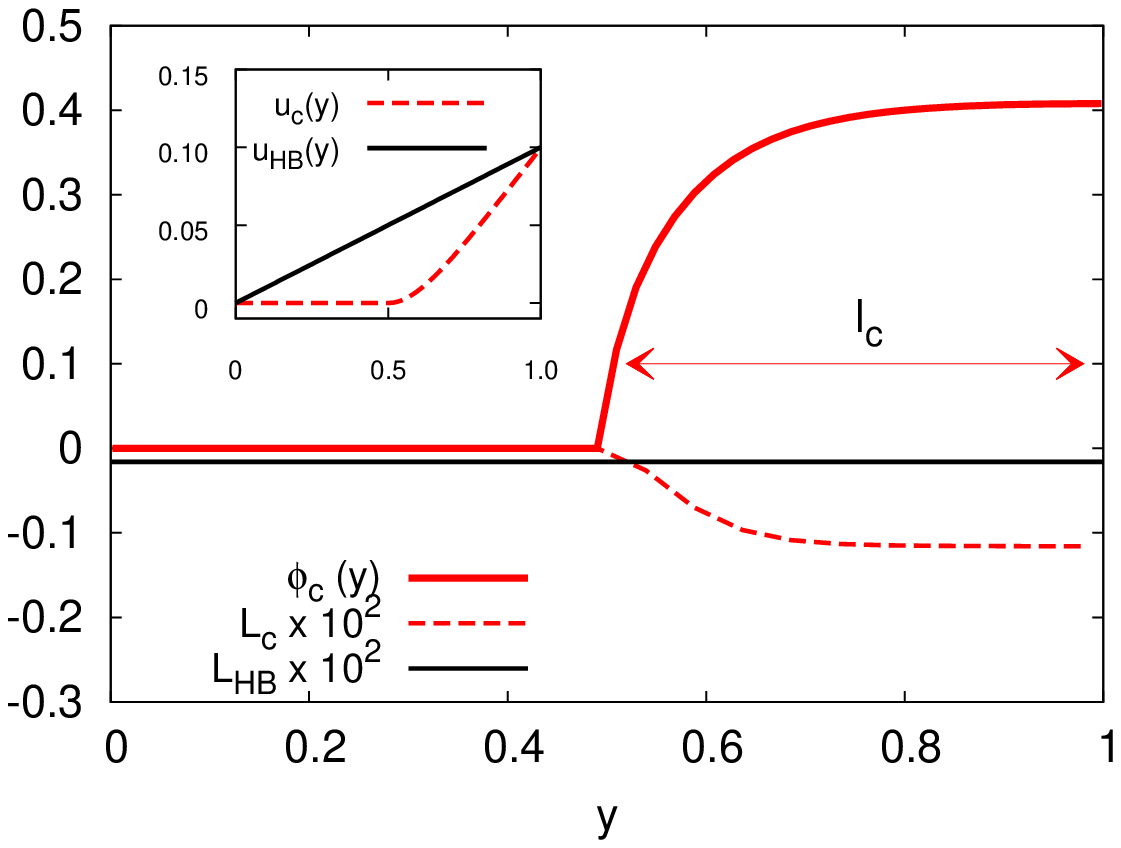}
\end{center}
\caption{Top panel: we show the global free-energy $F[\phi]$ (see equation \eqref{F}) for three different cases: linear velocity profile corresponding to a HB law (black solid line); compacton adjacent to the upper wall of the channel (see also bottom panel) with stress $\sigma=1.4$ (blue dot-dashed line) and $\sigma=1.5$ (red dashed line). At a given stress $\sigma$, the size of the compacton is changed in order to satisfy the constraint \eqref{vincolo}. The free-energy of the compactons is smaller than the linear velocity profile up to a critical shear where the size of the compacton becomes equal to the channel width. Bottom panel: free-energy densities ${L}(\phi,\partial_y \phi)$ defined in equation \eqref{F} as a function of the channel position at imposed shear $S=0.1$ (subscripts indicate the compact (c) and the Herschel-Bulkley (HB) solutions). The chosen compact solution (red solid line) corresponds to a stress $\sigma=1.5$. In the inset we report the corresponding velocity profiles, for both compact and HB solutions. \label{HB_versus_SB}}
\end{figure}

As discussed in the previous Section, the energy landscape of our system is characterized by various stationary solutions corresponding to both homogeneous (HB solution) and non-homogeneous (compact) solutions. Now, we study the corresponding free-energy of those solutions and the resulting velocity profiles. To that purpose we go back to the geometry of the Couette flow at imposed shear and consider for simplicity the case of a compact solution with a compacton adjacent to the upper wall of the channel with the property $\phi_E(y)>0$ for $l_c \le y\le 1$ and $\partial_y \phi=0$ at $y=1$ (see bottom panel of Figure \ref{HB_versus_SB}). This choice of boundary conditions makes it easy to discuss the case of compact solutions close to the boundaries, although it differs from the usual choice of Dirichelet boundary conditions, where a new parameter representing the wall fluidity~\cite{Mansard13,Mansard14,Picard01} is introduced. In the Couette flow with imposed shear $S$, the overall free-energy of both the compact and HB solutions depends on the shear $S$ as follows: given the stress $\sigma$, the full non linear solution of equation \eqref{phi0} must be found for a value of $m(\sigma)=(\sigma-1)/\sigma^{1/2}$ consistent
with \label{vincolobeginning}
\begin{equation}\label{vincolo}
\sigma \int_0^1 \phi^2(y)\, dy = S.
\end{equation}
For the HB solution, corresponding to $\phi=\phi_m=m$, the constraint \eqref{vincolo} imposes $\sigma=S/m^2$ so that the HB relation \eqref{eq:msigma} implies $m(\sigma(S))=\frac{S^{1/2}}{(1+S^{1/2})^{1/2}}$ and the corresponding free-energy is given by
\begin{equation}\label{eq:FE_HB}
F_{HB}(S) \equiv - \frac{1}{10} \frac{S^{5/2}}{(1+S^{1/2})^{5/2}}.
\end{equation}
The free-energy for the compact solution is different. Given the shear $S$, the free-energy is not uniquely determined, as multiple choices of $\sigma$ and $E$ are compatible with the constraint \eqref{vincolo}. This makes the free-energy dependent on both $\sigma$ and $S$: once these two parameters are fixed, we are able to determine the appropriate $E$ in \eqref{E1} that makes possible to derive the $\phi_E$ that satisfies the constraint \eqref{vincolo}. For a given stress $\sigma$, this obviously leads to a larger $l_c$ at increasing $S$, {\em i.e.,} the size of the compacton is increased to verify the constraint \eqref{vincolo} (see the bottom panel of Figure \ref{fig:compacton}). In the top panel of Figure \ref{HB_versus_SB} we show the free-energy $F_{HB}(S)$ of the HB solution and the free-energy of the compact solution $F_{c}(S,\sigma)$ as a function of the shear $S$. For the compact solution, we choose two different values of the stress, $\sigma=1.4$ and $\sigma=1.5$. We see that the cases with compact solutions show a free-energy smaller than the HB value up to a critical shear $S_{cr}$ which is stress dependent
\begin{equation}
F_{c}(S,\sigma) \le F_{HB}(S) \hspace{.2in} S \le S_{cr}(\sigma).
\end{equation}
In correspondence of $S_{cr}$, the size of the compact region becomes of the order of the channel size. Compact solutions in $y$ imply that the shear, according to the definition of the fluidity $f=\phi^2$, is different from zero only in a compact region. This suggests, by itself, an intriguing link to {\it shear-banding}, since it permits coexistence of (compact) flowing and non-flowing states within the same spatial flow configuration. This can be evinced from the bottom panel of Figure \ref{HB_versus_SB}, where we analyze a situation with imposed shear $S=0.1$ and stress $\sigma=1.5$. \\
Central to this free-energetic picture is the imposition of the global constraint \eqref{vincolo}, which is key to tip the free-energy balance in favor of compact solutions versus HB solutions, $F_c(\sigma,S) \le F_{HB}(S)$, in the proper range of the shear $S$, as reported in Figure \ref{HB_versus_SB}. We wish to observe that, had we imposed the stress $\sigma$, we would have obtained
\begin{equation}\label{F2}
F_c (\sigma) =   2 \int_{y_0}^{y_1} \left[-\frac{1}{4} m(\sigma) \phi^4 + \frac{1}{5} \phi^4 |\phi|+ \xi^2 \phi^2 (\partial_y \phi)^2 \right] d y \ge F_{HB}(\sigma)
\end{equation}
meaning that the compact solution $\phi_c$ has larger free-energy than the HB solution. In other words, by imposing the stress instead of the shear, the HB solution would always show up.


\section{Stability of Compact Solutions: Time Dynamics and Aging}\label{sec:aging}

\begin{figure}[t!]
\begin{center}
\includegraphics[width=0.6\textwidth]{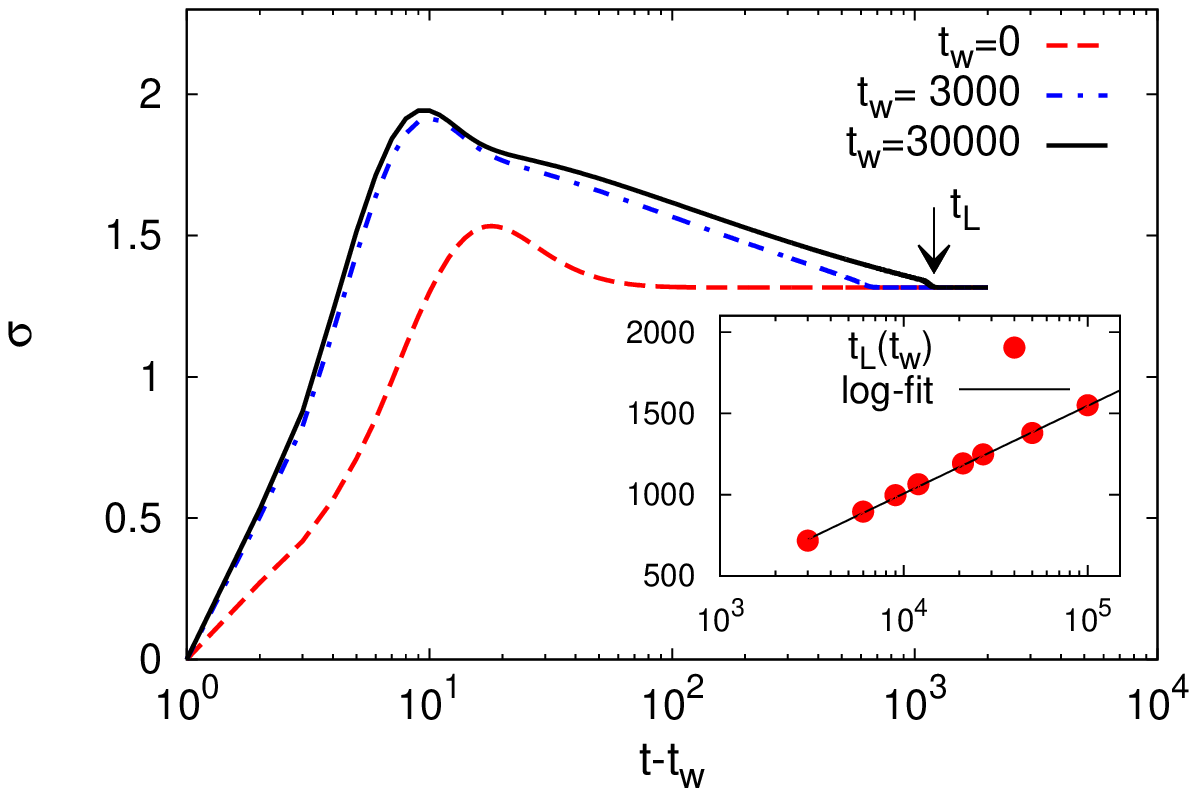}
\end{center}
\caption{The behavior of $\sigma(t,t_{\mathrm{w}})$ under the effect of the shear $S = 0.1$ applied to the system after the waiting time $t_{\mathrm{w}}$. The initial condition for $\phi$ are: $\phi=0.2$ for $y\in[0:0.8]$ and $\phi=0.5$ for $y\in [0.8,1]$. The black arrow indicates the time $t_L(t_{\mathrm{w}})$ at which the value of the stress reaches the HB value $1+S^{1/2}$. In the inset we show $t_L$ as a function of $t_{\mathrm{w}}$: a clear logarithmic dependence is observed. Details of numerical simulations are reported in the text (see Section \ref{sec:aging}).}
\label{fig:aging}
\end{figure}

The connection between shear-bandings and the compact solutions $\phi_c$ is extremely appealing, but we hasten to remark important features about the stability of the shear-banding solutions thus obtained. To this aim, we resort to the very simple equation
\begin{equation}\label{phi1}
\partial_t \phi = - \frac{\delta F}{\delta \phi}
\end{equation}
to define a time-evolution of the order parameter that is consistent with the attainment of local extrema of the free-energy functional in the long-time.  More sophisticated formulations of the dynamics may eventually be considered. \\
As a matter of fact, much of the insight conveyed by Figure \ref{HB_versus_SB} can also be gained by analyzing the scaling properties of the model. In particular, equations of motion \eqref{phi1} are invariant upon the scale transformation:
\begin{equation}\label{201}
y \rightarrow \lambda^a y \hspace{.1in} \phi \rightarrow \lambda^b
\phi \hspace{.1in} \xi \rightarrow \lambda^{b/2+a} \xi \hspace{.1in} t
\rightarrow \lambda^{-3b} t \hspace{.1in} m \rightarrow \lambda^b m.
\end{equation}
By using \eqref{201} we determine the scaling properties of the free-energy as
\begin{equation}
F \rightarrow \lambda^{5b+a} F.
\end{equation}
The choice $a=0$ and $b=1$, corresponding to a linear velocity profile with increasing amplitude $m$ and no shape change, implies a free-energy that scales as $F \sim A_{HB} \lambda^5$. The choice $a=1$ and $b=0$ corresponds to compact solutions, namely increasing size of the shear-band with no increase in amplitude $m$, and delivers $F \sim A_{c} \lambda$. Note that both constants $A_{HB}$ and $A_{c}$ are negative, so that the minimum corresponds to the larger absolute value. Finally, under the assumption that $S$ is sufficiently small, {\em i.e.,} $\sigma \approx 1$, equation \eqref{vincolo} delivers $S_{HB} \sim \lambda^2$ for $a=0$ and $b=1$ and $S_{c} \sim \lambda$ for $a=1$ and $b=0$. Based on the above scaling relations, we obtain:
\begin{equation}\label{eq:F_scaling1}
F_{HB} \sim A_{HB} S^{5/2}
\end{equation}
\begin{equation}\label{eq:F_scaling2}
F_{c} \sim A_{c} S.
\end{equation}
The result is that, up to a critical value $S=S_{cr} \equiv (A_{c}/A_{HB})^{2/3}$, one has $F_{c} \le F_{HB}$, so that the compactons are favored with respect to the homogeneous configurations. Scaling laws \eqref{eq:F_scaling1}-\eqref{eq:F_scaling2} correspond to the functions reported in Figure \ref{HB_versus_SB}. \\
Given the time dynamics \eqref{phi1}, one can verify that the compactons $\phi_E$ are stable against perturbations of the order parameter. In particular, one can compute $ \delta F \equiv F[\phi_E+\delta \phi]-F[\phi_E]$ up to the second order term in $\delta \phi$, and show that $\delta F$ is positive defined for an energy $E$ sufficiently close to $m^5(\sigma)/20$ (see Appendix \ref{appendix1}). However, the non-flowing state $\phi_0$ in \eqref{eq:compacton} is unstable, and the overall stability of the system must take into account both compact and vacuum components. The compacton is thus expected to grow and ``eat up'' the unstable non-flowing state $\phi_0$. By this process, a critical situation is attained whenever the compacton hits the size of the channel: at that point, the HB solution is energetically favored again and the compacton yields to an extended HB profile, going back to a situation with a global bulk rheology. Under such conditions, and consistently with the fact that compact regions evolve at a faster rate than the ``vacuum'', it is apparent that the time-relaxation of the overall compact+vacuum system becomes heterogeneous not only in space but also in time, {\em i.e.,} the systems shows aging. To highlight this effect we have conducted numerical simulations: the interval $[0,1]$ has been discretized with $512$ collocation points and an Euler-Cauchy scheme with integration step ${\mathrm{dt}}=10^{-3}$ has been used for the time dynamics of the free-energy. The parameter $\xi$ has been fixed to $\xi=0.04$. At time $t=0$, we start with an initial condition $\phi=0.2$ in the region $[0:0.8]$ and $\phi=0.5$ in $[0.8:1]$, {\em i.e.,} we start with a non-uniform initial fluidity. We integrate the equation of motion for a time $t_{\mathrm{w}}$ and, at $t=t_{\mathrm{w}}$ we apply a shear $S=0.04$ and we compute the resulting stress $\sigma(t,t_{\mathrm{w}})$. In Figure \ref{fig:aging}, we show $\sigma(t,t_{\mathrm{w}})$: after an almost linear growth, the stress reaches a maximum $\sigma_M(t_{\mathrm{w}})$ and then eventually decays to the HB value $1+S^{1/2}$ at time $t_L(t_{\mathrm{w}})$. Note that $\sigma_M$ and $t_L$ depend on the waiting time $t_{\mathrm{w}}$, in particular there is an overshoot that depends on the age $t_{\mathrm{w}}$ of the sample~\cite{Fielding3}. Similar results have been reported experimentally in~\cite{Manneville2011a} and in MD simulations of Lennard-Jones glasses~\cite{Varnik1,Varnik2} (see also~\cite{Fielding14} for a recent review). In the inset of Figure \ref{fig:aging}, we show $t_L$ as a function of $t_{\mathrm{w}}$. It is apparent how it is very well fitted by a logarithmic function of $t_{\mathrm{w}}$.  We remark that the same effect does not appear if we consider the dynamic equation for the fluidity as
\begin{equation}
\label{naive0}
\partial_t f = - \frac{\delta F}{\delta f}.
\end{equation}
In fact, upon multiplying equation \eqref{phi1} by $\phi$, we can
rewrite equation \eqref{phi1} in the form
\begin{equation}\label{naive1}
\partial_t f = -f \frac{\delta F}{\delta f}.
\end{equation}
The comparison between equations \eqref{naive0} and \eqref{naive1} highlights the physical meaning of the relation $f=\phi^2$: upon the assumption of a simple first order steepest-descent dynamics \eqref{phi1}, the transformation from $f$ to $\phi$ implies that the former evolves on a typical configuration-dependent time scale $1/f$ \eqref{naive1}, so that non-flowing states $f \sim 0$ take a virtually infinite time to relax, as opposed to the flowing ones. This configuration-dependent time-scale separation lies at the heart of the aging phenomena described in Figure \ref{fig:aging}.


\section{Effects of Mechanical noise at imposed Stress}\label{sec:noise}

\begin{figure}[t!]
\begin{center}
\includegraphics[width=0.6\textwidth]{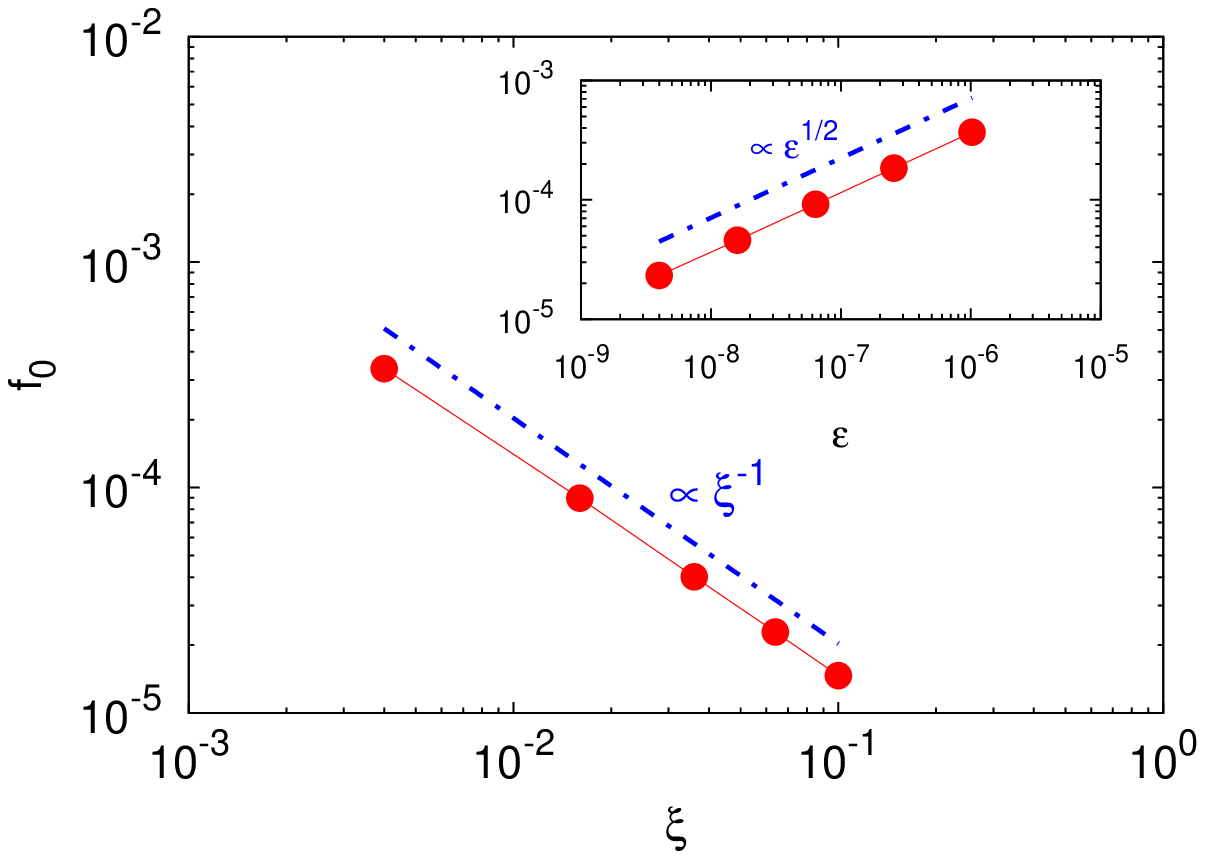}
\caption{We report the average fluidity $f_0=\langle \phi^2 \rangle$ (see also equation \eqref{f0}) obtained from numerical simulations of the model equations \eqref{eq:modelequation}. Main panel: the average fluidity as a function of $\xi$ for $\epsilon= 10^{-8}$. Inset: $f_0$ as a function of the noise strength $\epsilon$ for $\xi=0.04$. In both cases the scaling predictions of equation \eqref{f0} are verified. All simulations have been performed at constant stress $\sigma=1$ ({\em i.e.,} $m=0$ in \eqref{eq:modelequation}). Details of numerical simulations are reported in the text (see Section \ref{sec:aging}).}
\label{fig6}
\end{center}
\end{figure}

In the previous Section we have shown that in a Couette flow the coupling between the external shear and the stress is such that compactons $\phi_E$ are stable states and the associated free-energy $F_{c}(S,\sigma)$ is smaller than the HB value up to a critical shear $S_{cr}$. However, the region where the order parameter is zero (the non-fluidized band) is unstable, so that the formation of permanent shear-bands is not possible and they can be observed only for a finite (possibly long) time. So much for the deterministic picture. The next natural question, is to inspect the behavior of the system under stochastic perturbations (noise). We shall refrain from identifying such noise with any thermodynamic temperature~\cite{Sollich3,Bouchbinder1,Bouchbinder2,Bouchbinder3}. Actually, we rather think of it as a mechanical noise due to dynamic heterogeneities. A naive expectation about the effect of the noise is that it raises the instability of the background field $\phi_{0}$ so that the system reaches the HB solution in a shorter time. However, this expectation is not true and we will discover something unexpected, {\em i.e.,} that the vacuum solution - and hence the compact solutions - are stabilized by noise. For the sake of simplicity, we include the constraint \eqref{vincolo} in the next Section whereas, here, we analyze the role of fluctuations at constant stress. We consider the time evolution
\eqref{phi1}:
\begin{equation}\label{eq:modelequation}
\begin{cases} \partial_t \phi = - \frac{\delta F}{\delta \phi} + \sqrt{\epsilon} \mathrm{w}(y,t) \\
F[\phi]= 2 \int_0^1 \left[-\frac{1}{4} m(\sigma) \phi^4 + \frac{1}{5}
  \phi^4 |\phi|+ \xi^2 \phi^2 (\partial_y \phi)^2 \right] d y
\end{cases}
\end{equation}
at imposed external stress $\sigma$, {\em i.e.,} at imposed $m(\sigma)=(\sigma-1)/\sigma^{1/2}$. In the above $\mathrm{w}(y,t)$ is a $\delta$-correlated white noise in space and time
\begin{equation}
\langle \mathrm{w}(y_1,t_1) \mathrm{w}(y_2,t_2) \rangle =
\delta(y_1-y_2)\delta(t_1-t_2).
\end{equation}
It must be understood that the solution of equation \eqref{eq:modelequation} is defined with an ultraviolet cutoff $\lambda \equiv 1/k_M$ needed for the regularization at small scales. A first, non trivial, consequence of the time dynamics \eqref{eq:modelequation} is that for $m=0$ ({\em i.e.,} $\sigma < \sigma_Y$), the space averaged fluidity $f_0$ of the system does not vanish. As discussed in the Appendix \ref{appendix2}, we have
\begin{equation}\label{f0}
f_0 = \langle \phi^2 \rangle \sim \frac{ \sqrt{\epsilon k_M} } {\xi
  k_M}.
\end{equation}
These results can be tested quite accurately by using numerical simulations. In Figure \ref{fig6} we show the value of $f_0$ as a function of $\xi$ (main figure) for $\epsilon= 10^{-8}$, whereas in the inset of the same Figure we show $f_0$ as a function of $\epsilon$. In both cases, the scaling predicted by \eqref{f0} is extremely well verified supporting the theoretical results obtained in appendix \ref{appendix2}. With the interpretation of the fluidity as the rate of plastic events in the system, equation \eqref{f0} tells us that a non zero value of the fluidity is still present at zero external forcing ({\em i.e.,} $S=0$ and/or $\sigma < \sigma_Y$)~\cite{Durian08,Durian13,SoftMatter15}. Further discussions on the nature and properties of the noise will be reported in the end of the present Section and in the conclusions



\begin{figure}[t!]
\begin{center}
\includegraphics[width=0.60\textwidth]{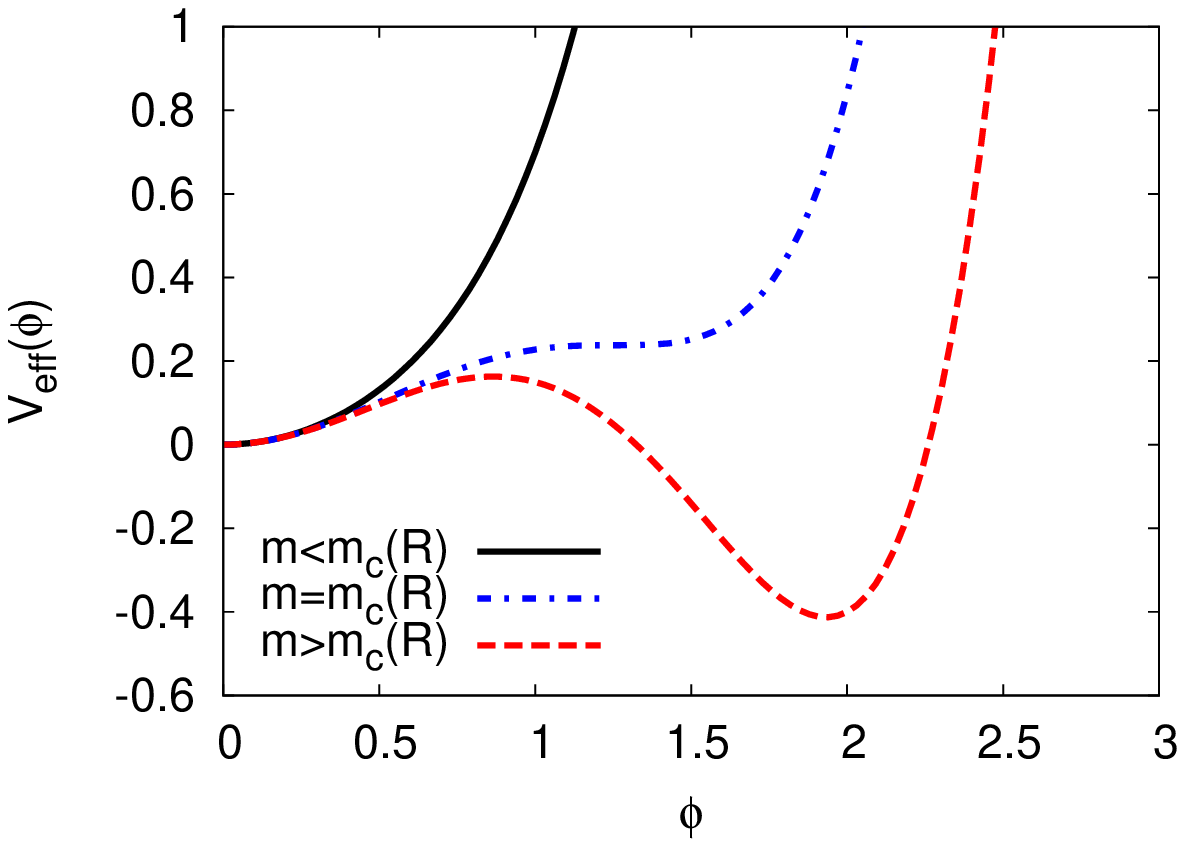}
\includegraphics[width=0.60\textwidth]{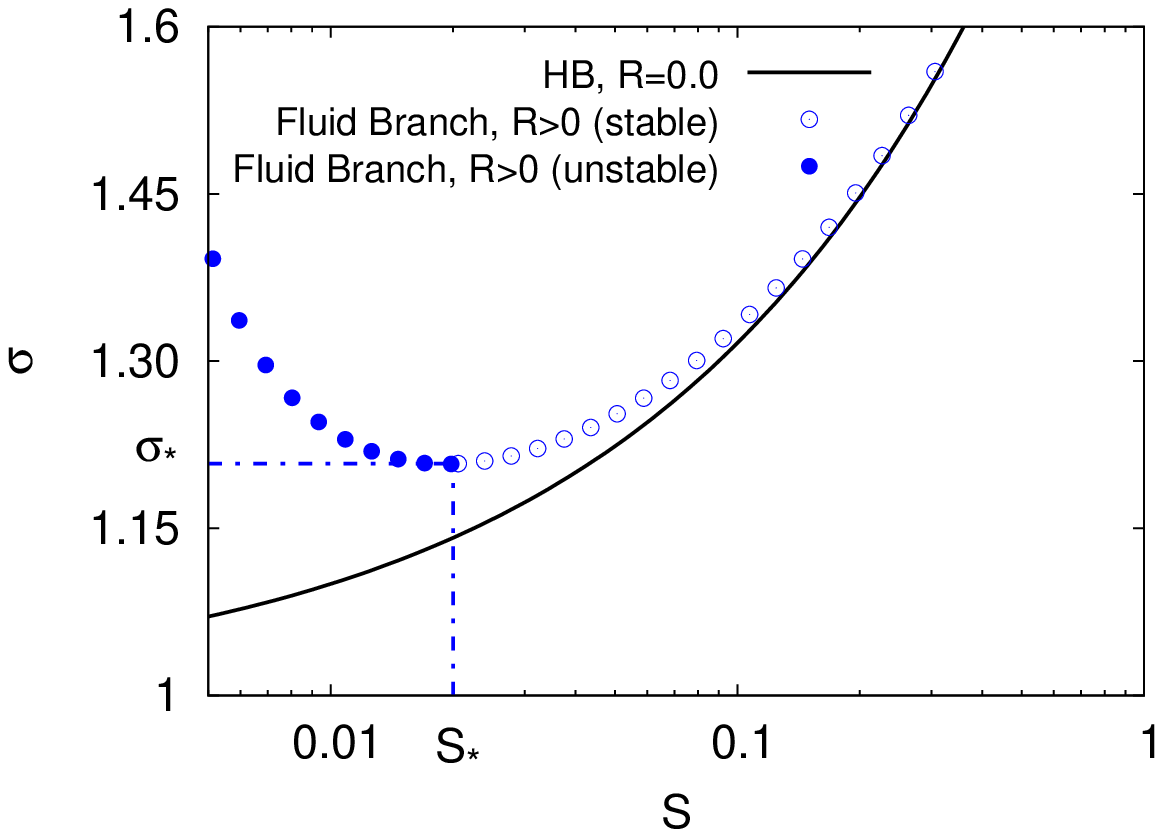}
\end{center}
\caption{Top panel: The effective potential $V_{\mbox{\tiny{eff}}}$ defined in equations \eqref{renorm} with $R=0.001$. A bifurcation is present at $m=m_*=(27R/4)^{1/3}$, above which two local minima appear at $\phi_{\mbox{\tiny{min}}}=\phi_0=0$ and at $\phi_{\mbox{\tiny{min}}}=\phi_m \sim m-R/m^2 + O((R/m^2)^2)$, together with a local maximum in $\phi=\phi_M$. Note that there is a range of $m$ ($m_* \le m \le m_c$) where the local minimum $\phi_m$ appears, but $\phi_0$ is still the global minimum. At the critical value $m=m_c$, the vacuum and homogeneous (stable) fluidized solution can coexist. Bottom Panel: homogeneous stationary flow-curve obtained from the local extrema equation $R-m \phi^2 +\phi^3=0$ (see text for details). Due to fluidity fluctuations ($R>0$), such flow-curve is a decreasing function up to the point in which it slopes upward at higher shear. This indicates the onset of an instability (see text for details).}
\label{fig:sketch}
\end{figure}

Next, we discuss the effect of the noise for $m>0$. Some interesting insights can be gained by using a self-consistent Hartree-like approximation~\cite{TarziaConiglio1,TarziaConiglio2,Chaikin}, as discussed in detail in Appendix \ref{appendix2}. This amounts to consider the free-energy
\begin{equation}\label{a7text}
F[\phi] = 2 \int_0^1\left[-{1 \over 4} m(\sigma) \phi^4 + {1 \over 5} \phi^4 |\phi | + \frac{R}{2} \phi^2 + \frac{D}{2} (\partial_y \phi)^2  \right]\,dy
\end{equation}
where
\begin{equation}
R \equiv 2 \xi^2 \langle (\partial_y \phi )^2 \rangle \hspace{.2in} D
\equiv 2 \xi^2 \langle \phi^2 \rangle
\end{equation}
need to be determined in a self-consistent way. Based on equation \eqref{eq:modelequation} and the results discussed in Appendix \ref{appendix2}, we can formally write the model equations as
\begin{equation}\label{renorm}
\begin{cases} \partial_t \phi =  - 2 \frac{d V_{\mbox{\tiny{eff}}}}{d \phi}+2 D \partial_{yy} \phi + \sqrt{\epsilon} \mathrm{w}(y,t) \\
V_{\mbox{\tiny{eff}}}(\phi) =  - \frac{1}{4} m(\sigma) \phi^4 + \frac{1}{5} \phi^4 | \phi | + \frac{R}{2}  \phi^2 .
\end{cases}
\end{equation}
Owing to renormalization effects, fluctuations turn the bare free-energy into an effective one, whose properties may lead to qualitatively new phenomena not contained in the original formulation. For the case in point, the potential $V$ flows into an effective one $V_{\mbox{\tiny{eff}}}$, supporting qualitatively new extrema through a renormalized ``mass'' term $\frac{R}{2} \phi^2$. Besides, a new diffusion term arises with no counterpart in the noise-free formulation. A full non-linear treatment would require that $R$ and $D$ were treated self-consistently, {\em i.e.,} taking into account their functional dependence on the configurational statistics of the system. However, as we shall see, significant insights can be gained by provisionally treating both quantities as constant parameters, and deferring a self-consistency check to a subsequent numerical solution. Thus, upon assuming $R$ and $D$ in \eqref{renorm} as ``constant'' parameters, we look
for local minima of the effective potential by solving ($\phi >0$):
\begin{equation}\label{a12text}
\begin{cases}
\left. \frac{d V_{\mbox{\tiny{eff}}}}{d \phi}
\right|_{\phi=\phi_{\mbox{\tiny{min}}}}=\left. \left(R\phi -m \phi^3 +
\phi^4\right) \right|_{\phi=\phi_{\mbox{\tiny{min}}}}=0 \\
\left. \frac{d^2 V_{\mbox{\tiny{eff}}}}{d \phi^2}
\right|_{\phi=\phi_{\mbox{\tiny{min}}}}>0.
\end{cases}
\end{equation}
It is easy to show that \eqref{a12text} has one solution $\phi_{\mbox{\tiny{min}}}=\phi_0=0$ for $m \le m_* \sim (27R/4)^{1/3}$, whereas for $m > m_*$, two local minima appear at $\phi_{\mbox{\tiny{min}}}=\phi_0=0$ and $\phi_{\mbox{\tiny{min}}}=\phi_m \sim m-R/m^2 + O((R/m^2)^2)$, together with a local maximum in $\phi=\phi_M$. This is shown in the top panel in Figure \ref{fig:sketch}. The local minimum $\phi_{m}$ is reminiscent of the HB solution already discussed in the previous Sections: this state receives a normalization and its stability changes, since now it appears as a local minimum only for $m \ge m_*$. Note that there is a range of $m$ ($m_* \le m \le m_c$) where the local minimum $\phi_m$ appears, but $\phi_0$ is still the global minimum. Above $m_c$ the minimum $\phi_m$ becomes the global minimum and the (stable) fluidized band is energetically favored. It is also interesting to extract the (stationary) homogeneous flow-curves from the local extrema solutions. Beyond the vacuum branch $\phi=\phi_0$, we obtain a fluidized branch from the equation $R-m \phi^2 +\phi^3=0$. In the bottom panel of figure \ref{fig:sketch} we show the stationary solutions, in terms of stress and shear, corresponding to this fluidized branch: unlike the simple HB model, in presence of fluctuations ($R>0$), this curve is a decreasing one before it slopes upward at higher shear. By fixing $m$ (i.e. the stress $\sigma$), the system is able to sustain the homogeneous fluid state only for $m \ge m_*$. For $m < m_*$, the only possible homogeneous solution is $\phi=\phi_0$. The decreasing part of the flow-curve corresponds to the local maximum $\phi_M$ of $V_{\mbox{\tiny{eff}}}$ and therefore, in terms of our dynamical equations, it is unstable. We remark that we impose the stress as a spatially constant variable and this instability is different from the ``mechanical'' instability induced by inertial terms~\cite{Picard01}. 






\begin{figure}[t!]
\begin{center}
\includegraphics[width=0.6\textwidth]{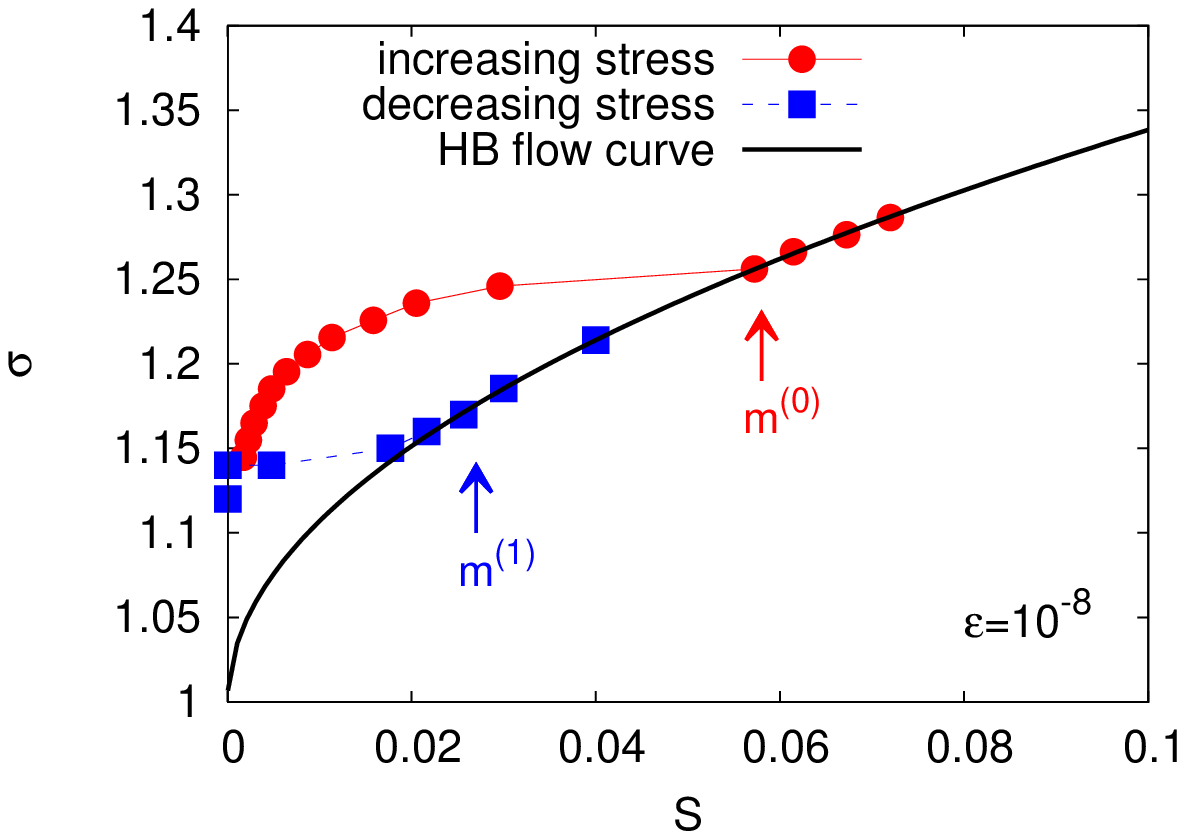}
\includegraphics[width=0.6\textwidth]{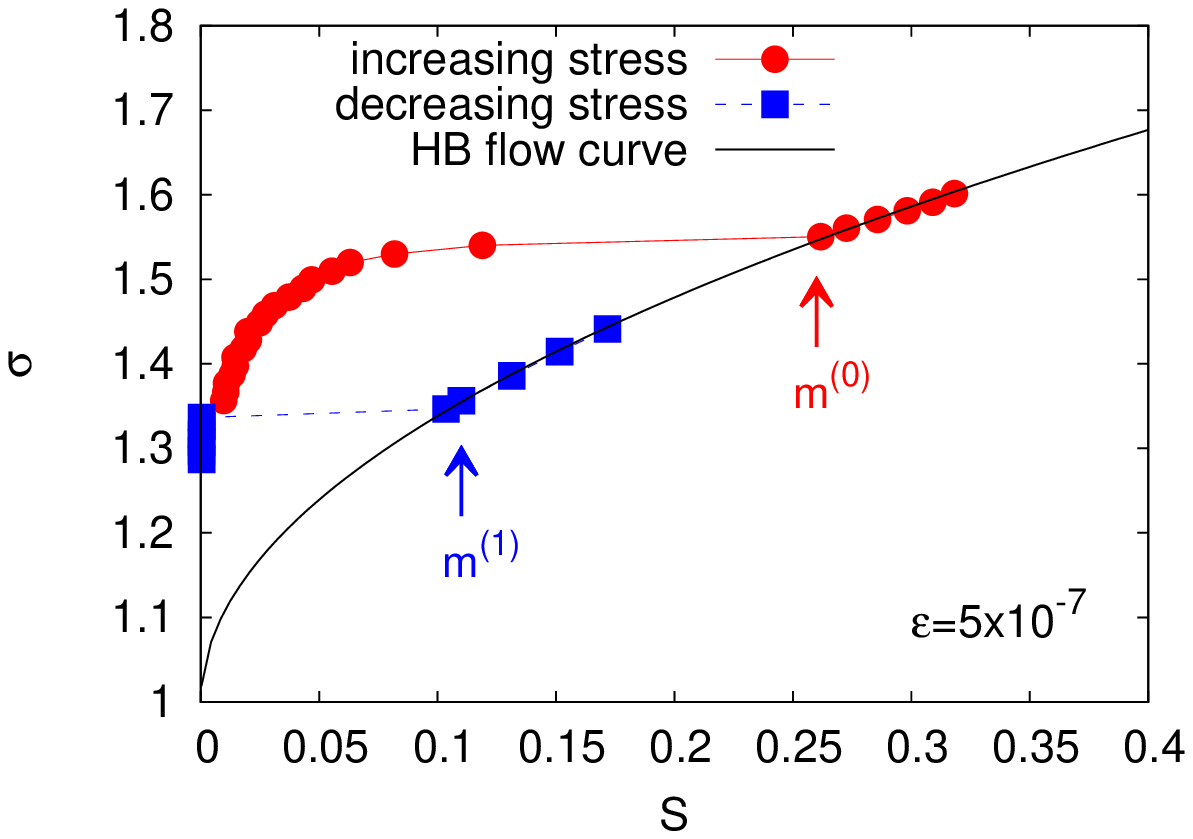}
\caption{Top panel: Stress shear relation obtained by the numerical simulations of \eqref{eq:modelequation} with $\epsilon = 10^{-8}$. Starting with an homogeneous initial condition $\phi=\phi_0=0$ we increase - step by step - the stress (red circles) up to the value where a clear HB behavior (solid line) is detected. Next we decrease the stress starting with $\phi=\phi_m=m$ (blue circles). A clear hysteresis cycle is observed. Arrows indicate the points where the values of $m^{(0)}( \epsilon )$ and $m^{(1)}(\epsilon)$ are extracted. Correspondingly, the critical values of the stresses can be obtained based on \eqref{eq:scaling0}-\eqref{eq:scaling1}. Bottom panel: same as the top panel with $\epsilon = 5 \times10^{-7}$.}
\label{fig:shear_stress}
\end{center}
\end{figure}

However, this is what we formally expect, based on the form of $V_{\mbox{\tiny{eff}}}$, by treating $R$ and $D$ as constant parameters. As mentioned earlier on, the complexity arises from the fact that both $R$ and $D$ depend on the fluctuations themselves \eqref{a12text}.  More importantly, $R$ and $D$ are {\it state dependent}, as can be seen by perturbative calculations ({\em i.e.,} assuming $\epsilon$ to be small) and by properly linearizing the potential near the local minima. In particular, an explicit computation (see Appendix \ref{appendix2}) shows that the scaling with respect to the noise strength $\epsilon$ is different, depending on the local minima ($\phi_0$ or $\phi_m$) around which they are computed:
\begin{equation}
R_0 \sim \xi k_M \sqrt{\epsilon k_M} \hspace{.2in} R_m \sim \epsilon
k_M/(2m^2).
\end{equation}
Consequently, the bifurcation point $m=m_c$ is different, depending on whether one linearizes around $\phi_0$ or $\phi_m$. We label these two bifurcation points $m^{(0)}$ and $m^{(1)}$, so defining also the corresponding critical stresses, $\sigma^{(0)}$ and $\sigma^{(1)}$, based on \eqref{eq:msigma}
\begin{equation}\label{eq:scaling0}
m^{(0)}(\epsilon) = \frac{\sigma^{(0)}-1}{(\sigma^{(0)})^{1/2}} \sim k_M^{1/2} \epsilon^{1/6}
\end{equation}
\begin{equation}\label{eq:scaling1}
m^{(1)}(\epsilon) = \frac{\sigma^{(1)}-1}{(\sigma^{(1)})^{1/2}} \sim k_M^{1/5} \epsilon^{1/5}.
\end{equation}
Due to the different scaling properties in $\epsilon$, we find that $m^{(1)} \le m^{(0)}$, or equivalently that $\sigma^{(1)} \le \sigma^{(0)}$. The above picture implies hysteresis in the system in the region $[\sigma^{(1)},\sigma^{(0)}]$. Indeed, any initial state close to $\phi_0$ is expected to attain the only stable solution of the system up to $m=m^{(0)}$, {\em i.e.,} up to a maximum stress $\sigma^{(0)}$. Similarly, starting with an initial state close to $\phi_m$, the system remains close to this state only for $m>m^{(1)}$, {\em i.e.,} only above a given stress $\sigma^{(1)}$. Therefore, once the stress falls within the range $\sigma \in [\sigma^{(1)},\sigma^{(0)}]$, two stable solutions are expected, depending on the initial conditions. As a matter of fact, these expectations are confirmed by the results of numerical simulations shown in Figure \ref{fig:shear_stress}. The system is initialized in the state $\phi_{0}$ and we then consider the time dynamics based on equation \eqref{eq:modelequation} with the noise in the range $[10^{-8},5\times 10^{-7}]$. We slowly change - step by step - the stress $\sigma$ by increasing $m$ and wait for the system to reach a stationary (in statistical sense) state, where we compute the apparent shear based on $S = \sigma \langle \phi^2 \rangle$. For this particular simulation, we must use as boundary condition $\phi=\phi_{\mathrm{w}}=m$. The reason for this choice is to avoid trapping in the stable state $\phi=\phi_0$ for any $m$. In the top panel of Figure \ref{fig:shear_stress}, we report the stress/shear relation obtained with $\epsilon=10^{-8}$: the red circles correspond to the simulations starting with the state $\phi_0$ and slowly increasing the stress; the blue squares correspond to the case where we start from state $\phi_m$ and a relatively large value of the stress, and then decrease the stress. In the bottom panel of Figure \ref{fig:shear_stress}, we show similar results obtained with the noise amplitude $\epsilon=5\times 10^{-7}$. Both figures support a well defined hysteresis effect, which is amplified by increasing the noise amplitude. Note that in the region $\sigma<\sigma^{(1)}$, only one stable solution is attained.

\begin{figure}[t!]
\begin{center}
\includegraphics[width=0.60\textwidth]{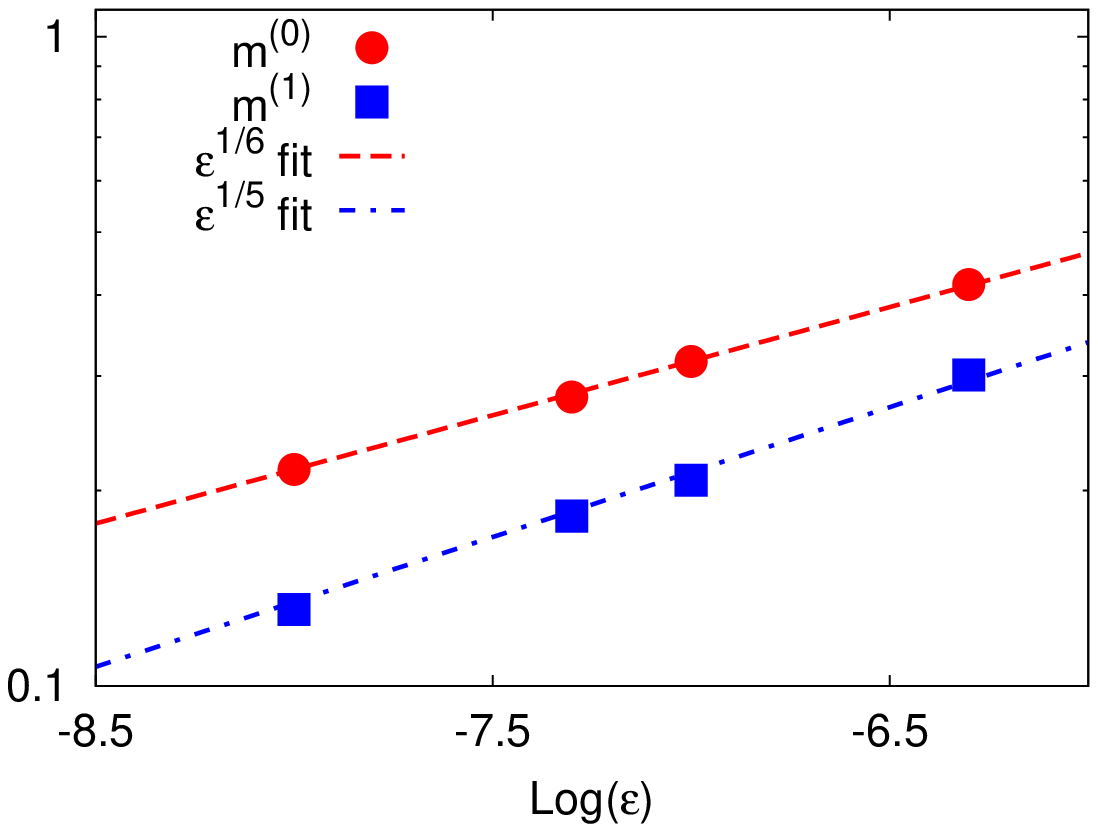}
\end{center}
\caption{We report the values of $m^{(0)}(\epsilon)$ (red circles) and $m^{(1)}(\epsilon)$ (blue squares) computed from the numerical integrations of \eqref{eq:modelequation} with noise amplitude in the range $[10^{-8}:5 \times 10^{-7}]$. The numerical procedure to determine the transition values, $m^{(0)}(\epsilon)$ and $m^{(1)}(\epsilon)$, is highlighted in Figure \ref{fig:shear_stress}.}
\label{fig:m1m2}
\end{figure}

From the shear-stress curves we can also determine the corresponding values of the stresses at which the transition occurs. These are reported in Figure \ref{fig:m1m2}, where we plot both $m^{(0)}$ (red circles) and $m^{(1)}$ (blue squares) as a function of $\epsilon$. The scaling predictions from equations \eqref{eq:scaling0}-\eqref{eq:scaling1} are also reported: note that although the two scaling laws are close to each other, it is clear that the numerical results cannot be fitted with the same scaling exponents in $\epsilon$. Our simulations therefore indicate that our analytical results \eqref{eq:scaling0}-\eqref{eq:scaling1} are in very good agreement with the complex non linear dynamics of the system. The existence of a hysteresis effect in the system is an important point since it is one of the many puzzling results obtained in laboratory experiments in soft glasses (see~\cite{Coussot02b,Becu,Paredes,Ovarlez08,Moller08} and references therein). Here, we can state that the hysteresis effect is due to the noise in the system and to the renormalization effects in the dynamics.


\begin{figure}[t!]
\begin{center}
\includegraphics[width=0.6\textwidth]{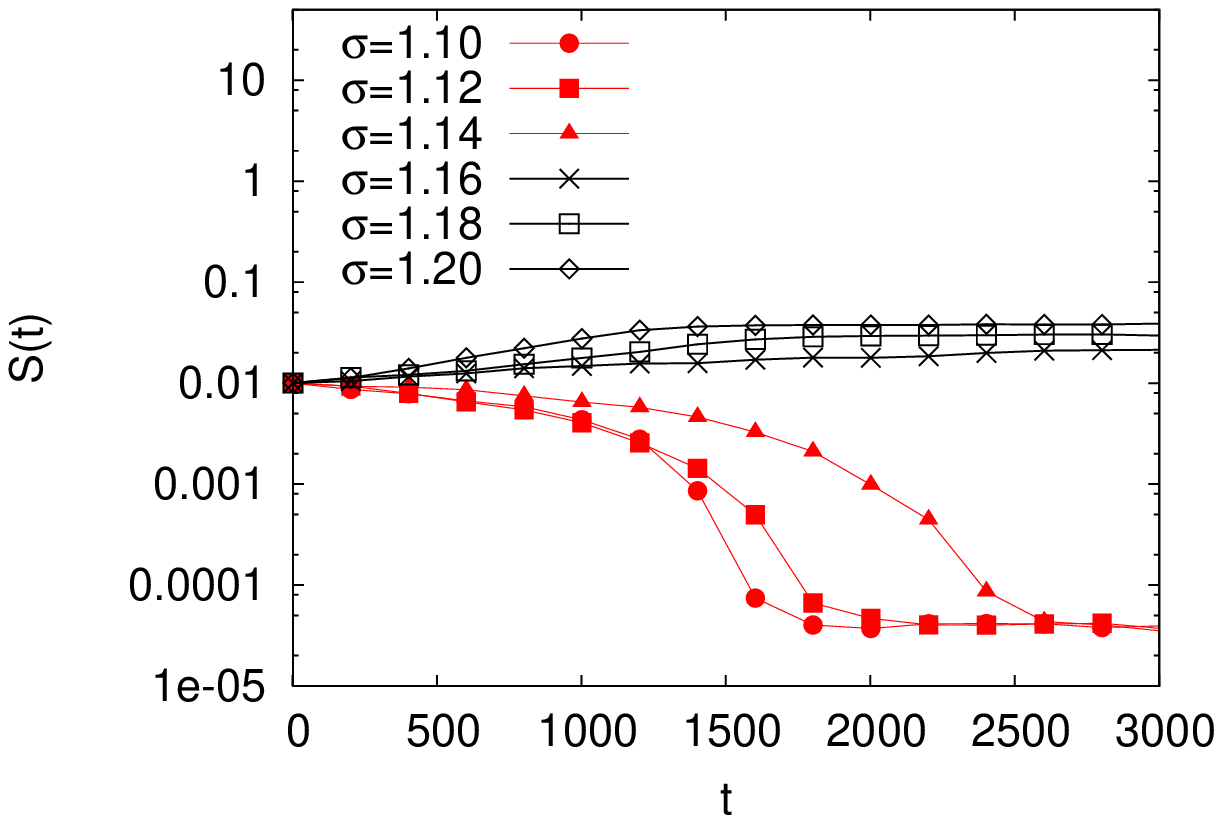}
\includegraphics[width=0.6\textwidth]{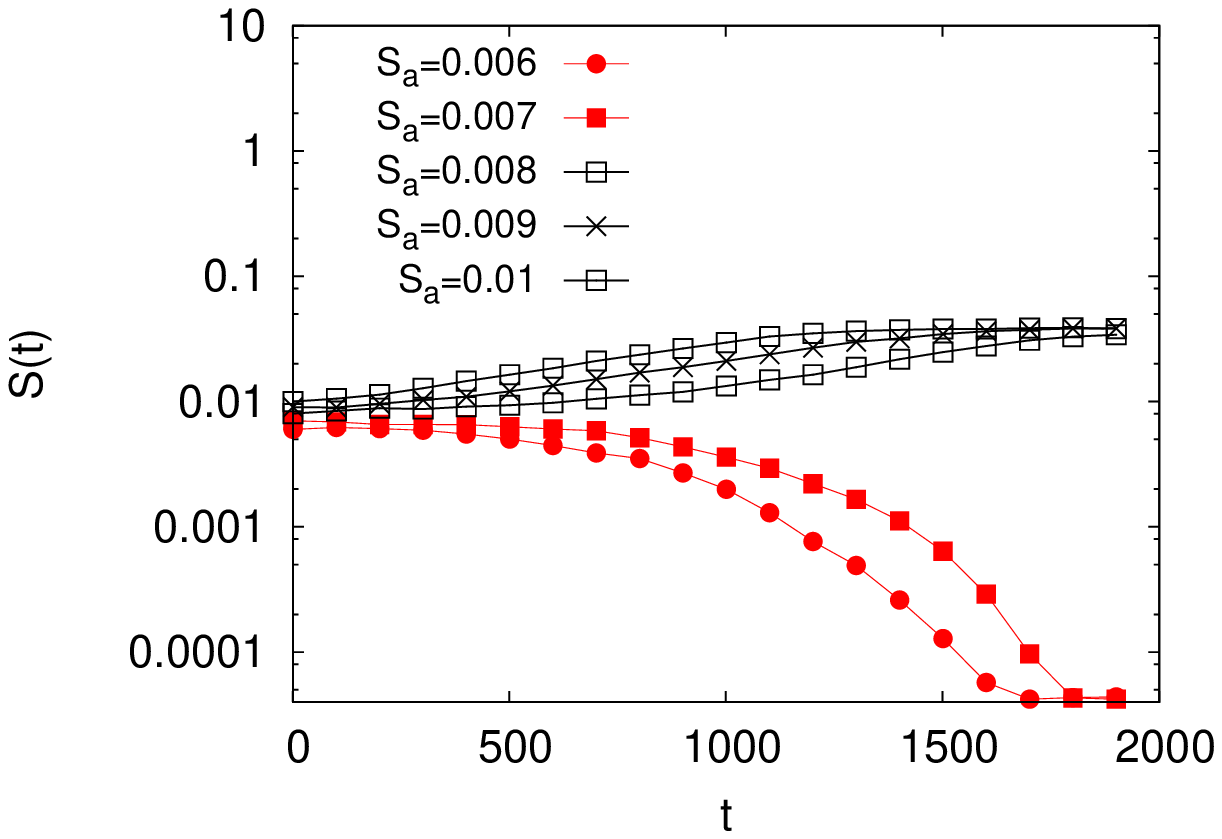}
\caption{Top panel: time dynamics of a pre-sheared state under the influence of a constant stress $\sigma$ based on equation \eqref{eq:modelequation}.  We consider an initial homogeneous state $\phi(y,0)=\phi_a$ corresponding to a pre-sheared state with shear $S(0)=S_a \equiv \phi_a^2 \sigma =0.01$. We then study the behavior of the apparent shear $S(t) = \sigma \int_0^1 \phi^2(y,t) \, dy$ for different values of the applied stress $\sigma$.  There is a clear bifurcation between the values of the stress below a critical value $\sigma^{(c)}$, at which the shear tends to vanish, and those above $\sigma^{(c)}$, where the shear goes asymptotically to a non zero value. Bottom panel: we show $S(t)$ for different values of the initial shear $S_a$ at fixed applied stress $\sigma=1.2$. For small values of $S_a$, the apparent shear tends to vanish whereas for $S_a$ greater than some critical value, the shear goes asymptotically to a non zero value. The numerical simulations are based on the model equations \eqref{eq:modelequation} with $\epsilon  = 10^{-8}$, $\xi=0.04$. \label{fig:bifurcation}}
\end{center}
\end{figure}

We can also use our findings to study other peculiar facets of the phenomenon of viscosity bifurcation. To this aim, we consider an initial pre-sheared state with fixed shear $S_a \equiv \phi_a^2 \sigma = 0.01$ at time $t=0$. This is realized with a homogeneous initial condition $\phi(y,0)=\phi_a=\sqrt{S_a/\sigma}$. We then vary the applied stress $\sigma$ and study the behavior of the apparent shear $S(t) = \sigma \int_0^1 \phi^2(y,t) \, dy$ as a function of time. In the top panel of Figure \ref{fig:bifurcation}, we show $S(t)$ for different values of the applied stress $\sigma$ with $\epsilon = 10^{-8}$, $\xi=0.04$. There is a clear bifurcation between the values of the stress below a critical value $\sigma^{(c)}$, at which the shear tends to vanish, and those above $\sigma^{(c)}$, where the shear goes asymptotically to a non zero value. Similarly, in the bottom panel of Figure \ref{fig:bifurcation}, we show $S(t)$ for different values of the initial shear $S_a$ with $\epsilon = 10^{-8}$, $\xi=0.04$ and fixed applied stress $\sigma=1.2$. For small values of $S_a$, the apparent shear tends to vanish whereas for $S_a$ greater than some critical value, the shear goes asymptotically to a non zero value.


Before closing this Section, we would like to remark that there is a simple argument to obtain the scaling laws reported in equations \eqref{eq:scaling0}-\eqref{eq:scaling1}. To this aim, let us consider the probability distribution $P[\phi]$ given by:
\begin{equation}\label{prob}
P[\phi] = {\cal Z}^{-1} \exp \left[ - \frac{1}{\epsilon} \int_{0}^{1}
  \left( - \frac{m}{4} \phi^4 + \frac{\phi^4}{5} |\phi|+\xi^2
  \phi^2 (\partial_y \phi)^2\right) \, dy \right].
\end{equation}
It is easy to verify that \eqref{prob} is invariant under the scale transformation
\begin{equation}\label{304}
y \rightarrow \lambda^a y \hspace{.1in} \phi \rightarrow \lambda^b
\phi \hspace{.1in} \xi \rightarrow \lambda^{b/2+a} \xi \hspace{.1in} m
\rightarrow \lambda^b m \hspace{.1in} \epsilon \rightarrow
\lambda^{5b+a} \epsilon.
\end{equation}
For $a=0$ the scale transformation \eqref{304} ignores spatial structures in the system and we obtain a scaling given by $m \sim \epsilon^{1/5}$. For $b=1$ the scale transformation \eqref{304} takes into account the spatial structure and we obtain the scaling $m \sim \epsilon^{1/6}$. Therefore the scale transformation \eqref{304} makes it possible to understand why there exists two different scaling regimes \eqref{eq:scaling0}-\eqref{eq:scaling1}. It is important to remark that the identification of $P$ as the probability distribution implies that the system is in equilibrium.\\
It is worthwhile to discuss in more detail the peculiar effects of the noise in our theory. As we already pointed out in the Introduction, we chose a very simple form of the noise to compute analytically the renormalization effects of the free energy. We remark that renormalization effects can be computed independently of a ``free energy formulation'' for the fluidity. In principle, the noise can depend on the external forcing and/or the rate of energy dissipation and it can be correlated in space and time. The effect of space/time correlations can be computed by a suitable (non trivial) generalization of the methods illustrated in Appendix \ref{appendix2}. Similarly, given a prescribed formulation of the noise variance as a function of the forcing, it is possible to generalize our results. At the present stage, we wish to keep our discussion as simple as possible. We argue that the qualitative picture emerging from our results is quite robust and independent (qualitatively) of the structure of the noise.

\section{Effects of Mechanical Noise at Imposed Shear}\label{sec:couette}

\begin{figure}[t!]
\begin{center}
\includegraphics[width=0.6\textwidth]{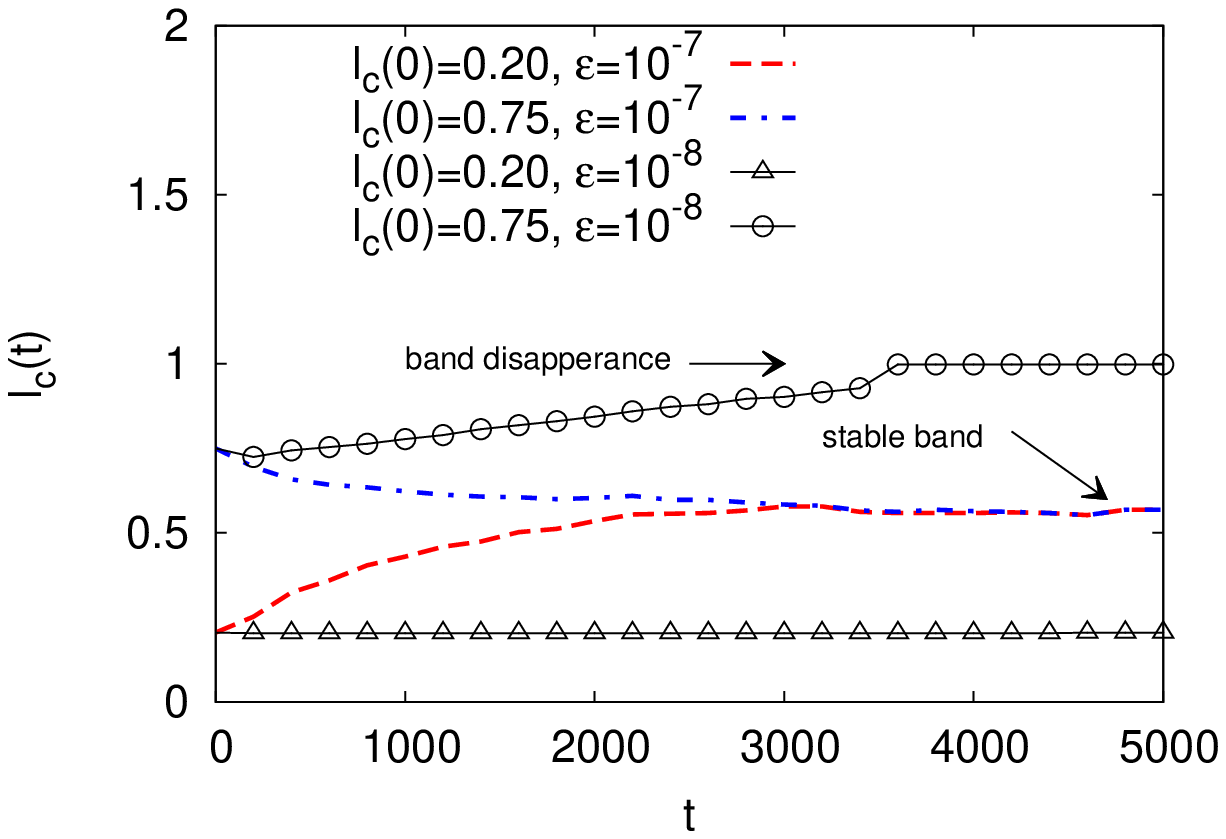}
\includegraphics[width=0.6\textwidth]{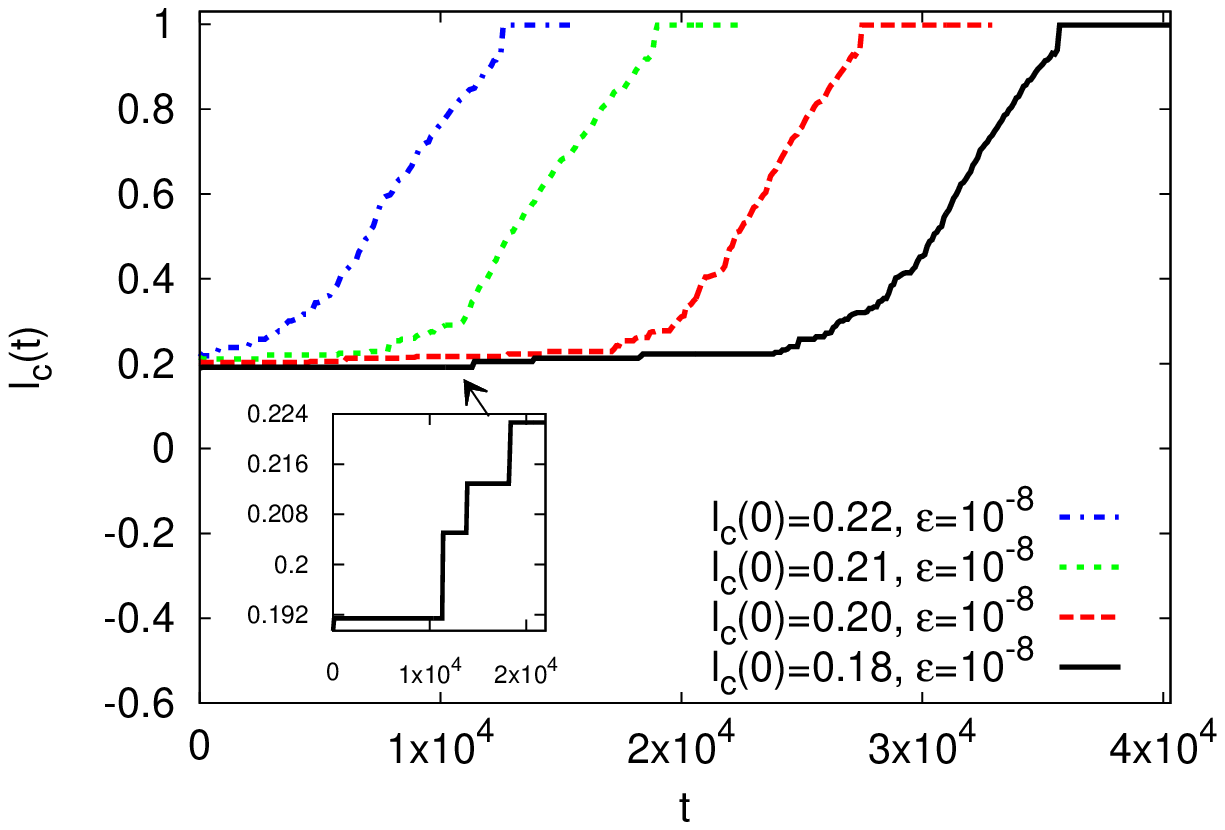}
\end{center}
\caption{Top panel: we show the band size $l_c(t)$ as a function of time in a Couette flow with two different initial conditions, $l_c(0)=0.20$ and $l_c(0)=0.75$, and $2$ different noise strengths, $\epsilon=10^{-7}$ and $\epsilon=10^{-8}$. Bottom panel: we report the band size $l_c(t)$ as a function of time with initial conditions chosen in the range $[0.18:0.22]$ and noise strength $\epsilon=10^{-8}$. The inset reports the early stages of the time dynamics for the case with $l_c(0)=0.18$, showing the increase of the band through a sequence of steps. In all cases the imposed shear is $S=0.04$.}
\label{fig:bands_time}
\end{figure}

\begin{figure}[t!]
\includegraphics[width=0.6\textwidth]{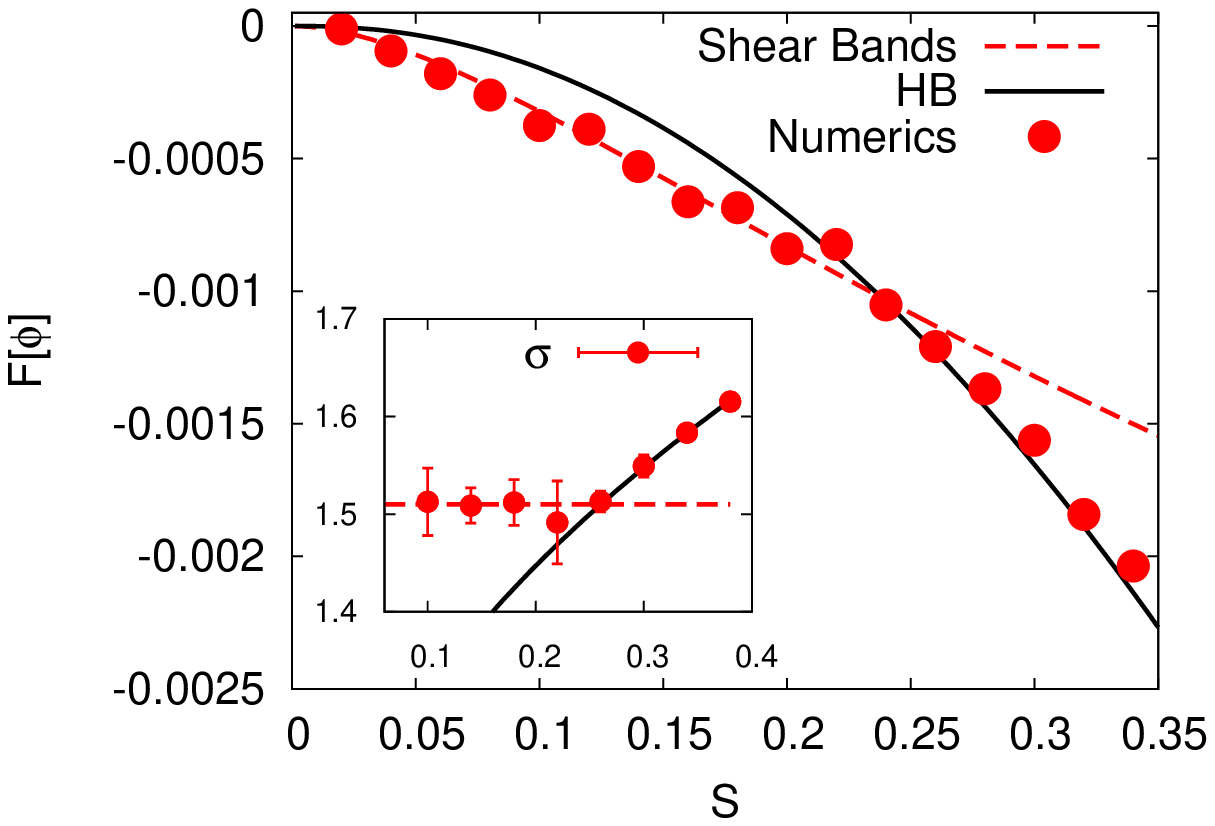}
\includegraphics[width=0.6\textwidth]{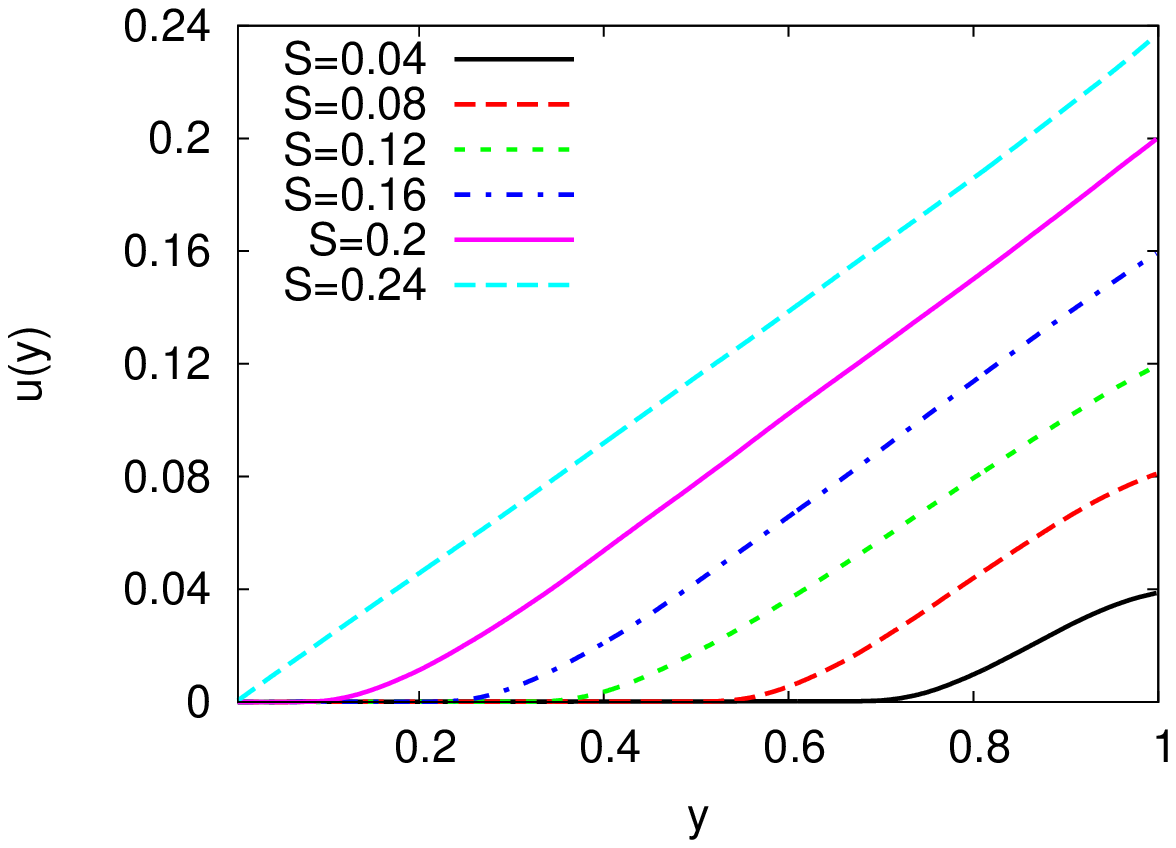}
\caption{Top panel: we show the free-energy for a Couette flow as a function of the apparent imposed shear $S$ and for $\epsilon=5\times 10^{-7}$. The red bullets correspond to the space-averaged free-energy $F[\phi]$ in the stationary state obtained from numerical simulations of the model equations \eqref{eq:modelequation}; the red dashed line is the value of $F$ for compact solutions \eqref{eq:F_scaling2}; the black solid line is the value of $F$ corresponding to the linear HB velocity profile \eqref{eq:FE_HB}. In the inset we show the stress/shear relation obtained by the numerical simulations. A clear plateau at $\sigma \approx 1.5 $ is observed corresponding to the value $\sigma^{(0)}$ of equation \eqref{eq:scaling0}. Bottom panel: the velocity profiles corresponding to the shear-band region of the top panel are reported.}
\label{fig:finalplot}
\end{figure}

Given the findings reported in Section \ref{sec:noise}, we now consider the dynamics given by equation \eqref{eq:modelequation}, together with the constraint \eqref{vincolo} in a Couette flow. If the vacuum $\phi_0$ is stabilized by fluctuations, the overall stability of compact solutions (hence the co-existence of flowing and non-flowing stable states) discloses the possibility to support stable shear-band configurations. Based on the results reported in Section \ref{sec:noise}, we expect the following scenario: the vacuum branch and the stable fluidized branch are separated by a range of shear. By fixing the imposed shear in such a range, a (stationary) homogeneous solution would be unstable, as it would fall in the decreasing part of the flow-curve (see figure \ref{fig:sketch}). We also know that compactons $\phi_E$ correspond to a minimum of the free-energy, a property which stays unchanged even in presence of external noise. If so, the compact solution (vacuum + compacton) could then be chosen.
Even in case the homogeneous fluidized solution becomes stable, the compacton solution would still be selected as long as it manages to attain a lower value of the free energy. These expectations are indeed borne out by numerical simulations that highlight the way in which the stress and the size of the compacton adjust matching the imposed shear. Let us first consider the dynamics of the system at an imposed shear $S=0.04$ with ``weak" and ``sufficiently strong" noise, $\epsilon=10^{-8}$ and $10^{-7}$. By ``strong'' we imply fluidity fluctuations in the order of a few percent of the deterministic value $f = m$, whereas fluctuations induced by a ``weak'' noise are below such threshold. For each noise strength, we consider two initial band sizes below and above the half size of the channel, $l_c(0)=0.2$ and $l_c(0)=0.75$. Using \eqref{eq:scaling0} and the analysis presented in the previous Section, we estimate that $\sigma^{(0)}(\epsilon = 10^{-8}) \approx 1.2$ and $\sigma^{(0)}(\epsilon = 10^{-7}) \approx 1.4$.  The numerical results obtained using the four possible combinations (initial conditions and noise strengths) are shown in the top panel of Figure \ref{fig:bands_time}, where we plot the time evolution of the shear-band $l_c(t)$. Let us begin by discussing the results with ``strong'' noise, $\epsilon = 10^{-7}$. With $l_c(0)=0.75$, the initial stress is below $\sigma^{(0)}(\epsilon = 10^{-7}) \approx 1.4$. In order to minimize the global free-energy (see also Figure \ref{HB_versus_SB}), the system {\it increases} the stress, {\em i.e.,} it {\it decreases} the size of the shear-band $l_c$, based on \eqref{vincolo}. This shrinking process goes on until $l_c$ becomes short enough to take $\sigma$ close to $\sigma^{(0)}$. At this stage, any further narrowing of the compacton would take $\sigma$ beyond the threshold $\sigma^{(0)}$, thus destabilizing the region $\phi=0$. Under such conditions, the shear-band solution increases its size again to $l_c$. In other words, $l_c$ is pinned down as the size compatible with the stress $\sigma=\sigma^{(0)}$, based on \eqref{vincolo}. The same reasoning carries on to the case $l_c(0)=0.20$, in which case $\sigma > \sigma^{(0)}$, so that $l_c$ increases until $\sigma$ reaches the value $\sigma^{(0)}$. Either ways, the only stable solution of the system corresponds to a shear-band with $\sigma=\sigma^{(0)}$. Let us now turn to the ``weak-noise'' scenario, $\epsilon=10^{-8}$. In this case, no shear-band solution is expected because $\sigma^{(0)} \approx 1.2$: the shear $S=0.04$ would select a point on the curve $F_c(\sigma=\sigma^{(0)},S)$ whose free-energy becomes comparable with the one of the HB solution, $F_c(\sigma=\sigma^{(0)},0.04) \approx F_{HB}(0.04)$, and the corresponding homogeneous solution is stable. Indeed, from Figure \ref{fig:bands_time}, we observe that, starting with $l_c(0)=0.75$, the system evolves towards metastable shear-bands, which disappear as soon as the compacton hits the size of the domain, after about $3000-4000$ time units. The case $l_c(0)=0.20$ presents however a different scenario. In this case, the numerical simulations show that $l_c$ remains basically constant, at least in the time span covered by the simulations. The reasons is that, even though that band is unstable, one must wait for a longer time before reaching the state $l_c=1$. Upon decreasing $l_c(0)$, the value of the initial stress $\sigma$ due to \eqref{vincolo} increases, and so it does the energy barrier protecting the asymptotic state $l_c=1$, leading to an even longer waiting time before the band is destabilized. This is appreciated from the bottom panel of Figure \ref{fig:bands_time}, where we observe that, in the initial stage, $l_c$ increases through a sequence of steps, each signaling the crossing of a corresponding barrier. Since the energy barrier decreases at increasing $l_c$, the overall process speeds up as time unfolds. Summarizing, for $\epsilon$ large enough to guarantee that the equilibrium probability distribution is given by \eqref{prob}, the non linear effect due to \eqref{vincolo} and the metastability of the region $\phi=0$, conspire to drive the system to a stable band compatible with a value of the stress $\sigma = \sigma^{(0)}(\epsilon)$. Thus, upon increasing the shear, we should observe a linear decrease of the free-energy with the shear $S$ (see equation \eqref{eq:F_scaling2}), up to the point when the shear-band size $l_c$ becomes of the order of the system size and the stable homogeneous solution is energetically favored again. This is again confirmed by the numerical simulations: in the top panel of Figure \ref{fig:finalplot} we show the value of the free-energy for a Couette flow as a function of the apparent imposed shear $S$. We compare the space-averaged free-energy evaluated from the numerical simulations (red bullets), the free-energy for compact solutions \eqref{eq:F_scaling2} (red dashed line), and the free-energy of the linear HB velocity profile \eqref{eq:FE_HB} (black solid line). The value of the noise amplitude is $\epsilon=5 \times 10^{-7}$. In the inset of the same figure, we show the stress/shear relation obtained through the numerical simulations. A clear plateau at $\sigma \approx 1.5 $ is observed, corresponding to the value $\sigma^{(0)}$ given in \eqref{eq:scaling0}. The velocity profiles are reported in the bottom panel of Figure \ref{fig:finalplot}.\\
Figure \ref{fig:finalplot} also demonstrates that the model defined by equations \eqref{vincolo}-\eqref{eq:modelequation} shows {\it stable} shear-bands for $\sigma \le \sigma^{(\mbox{\tiny{U}})}_{Y}$, where $\sigma^{(\mbox{\tiny{U}})}_{Y}$ is the value of stress  in the upper plateau of the shear-stress relation, {\em i.e.,} basically $\sigma^{(0)}$ in \eqref{eq:scaling0}. In other words, the stability of shear-bands as a local minimum of the functional $F[\phi]$ and the noise amplitude $\epsilon$ cooperate in such a way as to increase the value of the yield stress at which $\phi_0$ becomes unstable. This cooperative effect highlights the subtle and somewhat counterintuitive role of the mechanical noise. \\
Before closing this Section, a few comments on the role of $\xi$ are in order. All of the above results have been obtained using the same value $\xi=0.04$ and it is natural to ask how the picture discussed so far would change upon changing $\xi$. According to \eqref{eq:scaling0}, one would naively expect that by increasing $\xi$ the value of $\sigma^{(0)}$ should also increase. However, a subtler analysis reveals that this is not the case. Since $R \sim \xi^2 \langle (\partial_y \phi )^2 \rangle $, there are two competing mechanisms which concur to fix the value of $m^{(0)} \sim R^{1/3}$: the prefactor $\xi^2$ and the spatial average inherent to the definition of $R$. The former obviously increases with $\xi$, the latter however has just the opposite effect. Indeed, since the system is split in two regions, the state $\phi_0$ and the compacton $\phi_E$, the spatial average involves a factor $(1-l_c)$ which is apparently decreasing upon increasing $l_c$ via an increase of $\xi$. As a result, the system develops a much weaker dependence on $\xi$ than one would expect.

\section{Discussion}\label{sec:discussion}

The formation of shear-bands has been investigated by several authors, using a variety of different models~\cite{Boukany10,Ravindranath08,Tapadia06,Boukany08,Britton97,Helgeson09,Salmon03,STZ,Dhont,Fielding1,Mansard11,Martens12,Coussot02c,Coussot10,Fielding14,Jagla07,Jagla10,Picard01}. The permanent banding observed for viscosity-bifurcating YSF's echoes the shear-banding effects that have been extensively studied in polymers and wormlike micelles~\cite{Boukany10,Ravindranath08,Tapadia06,Boukany08,Britton97,Helgeson09,Salmon03}. There, the criterion for the formation of steady state shear-bands hinges on a non-monotonic shear-stress constitutive relation, whereby the negative slope would trigger the instability leading to heterogeneous flows and shear-bands out of an underlying homogeneous shear. Mesoscopic models based on a non-monotonic constitutive relation have also been put forward to explain shear-bands in viscosity-bifurcating YSF's~\cite{Fielding1,Mansard11,Martens12,Coussot02c,Coussot10,Picard01}, with the non-monotonicity that originates from different mechanisms. The well-known SGR (Soft Glassy Rheology) model proposed by Sollich and coworkers~\cite{Sollich1,Sollich2,Sollich3,Fielding1} describes the jamming transition in terms of trap dynamics for mesoscopic stress elements. Yield processes are taken as activated by some effective temperature $x$, based on the idea that yield events elsewhere in the material cause ``kicks'' which add up to an effective thermal noise~\cite{Sollich1,Sollich2,Sollich3}. The original SGR model captures a non-linear constitutive curve, yet a monotonic one, and thus fails to tie up with the shear-band instability mechanism discussed above.  Shear-bands can however be recovered by lifting the assumption of a constant effective temperature $x$, and considering a relaxation-diffusion dynamics for such parameter. For full details see the original paper by Fielding {\it et al.}~\cite{Fielding1}. An alternative model, by Mansard {\it et al.}~\cite{Mansard11}, resorts to the KEP description~\cite{Bocquet09}, supplemented with a phenomenological equation, to account for the coupling between the flow and the structure. The authors show that viscosity bifurcation occurs, due to a flow-induced weakening of the structure. Similar ideas, were studied by Dahmen and co-workers~\cite{Dahmen09}. Another approach, due to Martens {\it et. al.}~\cite{Martens12}, uses a minimalist mesoscopic model incorporating the local dynamics of plastic events (small-scale rearrangements, followed by stress redistribution). The authors studied the effects of the typical restructuring time needed to regain the original structure after a local rearrangement~\cite{Coussot10}. The spontaneous formation of permanent shear-bands is observed when this restructuring time is large compared to the typical stress release time. The formation of shear-bands due to structural relaxation has also been addressed in a detailed mesoscopic study by Jagla~\cite{Jagla07,Jagla10}. In the fluidity models considered by Picard {\it et. al.}~\cite{Picard01}, the authors study the relaxation dynamics of the fluidity via a free-energy functional approach, which includes ad-hoc deterministic terms to account for the competition between relaxation (aging) and flow-induced rejuvenation. Such competition gives rise to stationary solutions as local minima (maxima) of the free-energy functional, thus leading to stable (unstable) branches. To be noted that the authors impose the stress as a spatially constant variable, thus suppressing the ``mechanical'' instability induced by inertial terms. Destabilization of states is however triggered by local stability analysis of the stationary free-energy solutions. Recent studies also show that shear-banding can arise quite generically under time-dependent flow protocols~\cite{Moorcroft11,Moorcroft13}, even for materials with monotonic flow-curves. Once formed, these bands may persist only transiently, or may remain to a steady state, depending on whether the material's flow-curve is monotonic or not. Further studies~\cite{Varnik1,Varnik2,Chaudhuri} point to the existence of a static yield stress exceeding the dynamical yield stress, with the latter measured upon decreasing the shear down to zero: this produces a step discontinuity in the material's constitutive properties at zero shear and allows again the coexistence between unsheared and flowing bands. Yet another possible mechanism for genuinely steady state shear-banding without a non-monotonic constitutive curve, is the coupling between flow and concentration~\cite{Schmitt}. Such mechanism has been explored in the context of steady shear-banded states of hard sphere colloidal glasses in Besseling {\it et al.}~\cite{besseling10}.\\
The present work inscribes into the mesoscopic framework~\cite{Sollich1,Sollich2,Sollich3,Fielding1,Fielding2,Langer,Pouliquen,Derec,Picard01,Picard,Mansard11,Nicolas13,Mansard13} and hinges on the idea of cooperativity~\cite{Goyon08,Bocquet09,Goyon10,Geraud13}. All the results presented here stem from a free-energy picture (in analogy to equilibrium dynamics), a scenario that fits well within the framework of the simple models discussed by Picard {\it et. al.}~\cite{Picard01}.  More precisely, the free-energy functional \eqref{Ffm} has been designed to lead (upon minimization) to the HB and fluidity results given in~\cite{Goyon08,Bocquet09}, i.e., the ``mean-field'' scenario for the model. The built-in HB monotonic flow-curve does not admit permanent banding under conditions of a steady applied shear flow. However, the main point of the present work is that the addition of noise to the mean field scenario \eqref{eq:modelequation} proves instrumental in promoting a regime of instability with negative slopes in the flow-curve for homogeneous flows (see Section \ref{sec:noise}). We emphasize that also in other models described before~\cite{Fielding1,Martens12}, permanent bands could be obtained only by extending the investigation beyond the mean field level. The introduction of fluctuations on top of the cooperativity scenario has the effect of decorrelating fluidity fluctuations on scales shorter than the cooperativity length $\xi$, provided the order parameter is sufficiently small. Thus, fluctuations decrease the average fluidity, a mechanism which competes with the non-locality triggered by cooperativity effects (non-local squared gradient terms in \eqref{Ffm}) and conspires to form stable bands. Our bands may be viewed as a ``collection'' of plastic events within a solid region that remains elastic (vacuum), thus echoing the works by Martens {\it et al.}~\cite{Martens12} and Mansard {\it et al.}~\cite{Mansard11}, without assuming neither a direct model of the restructuring time~\cite{Martens12}, nor an equation describing the flow-structure coupling in the fluid~\cite{Mansard11}.\\
We also wish to emphasize the role played by diffusive terms. First of all, we remark that many rheological models have introduced diffusivity effects in the constitutive equations: in some models (see ~\cite{Fielding07} for a review) the diffusive terms are added to the stress equation and are needed to confer a finite interfacial width between the bands; other rheological models for shear-bands~\cite{Dhont} propose an expansion of the stress tensor with the inclusion of second order derivatives of the shear, referred to as the ``shear-curvature viscosity''. At variance with these models, we do not introduce any diffusivity effect in the stress, which is considered spatially constant~\cite{Picard01}. Nevertheless, we underline the crucial role played by diffusive effects in our model, which physically represent the spatial range of plastic relaxation. This effect is an essential mechanism to obtain heterogeneous compact solutions, as outlined in Section \eqref{sec:model}. Also, in our case, the diffusivity term is rather peculiar, see \eqref{naive1}. In particular, for small $f$ the effect of diffusion decreases and eventually vanishes at $f=0$. This is not the case of other models~\cite{Picard01}. We also emphasize that the negative slopes in the flow-curve result from the combined effect of both noise and cooperativity: failing either of these two, the renormalization of the flow-curve would be lost and no permanent shear-bandings would be observed.\\
We finally wish to stress that our model falls in the general arena of mesoscopic models~\cite{Martens12,Nicolas13,Nicolas13b} that do {\it not} include dilational effects which might take place during plastic events. These effects may well add corrections to the picture drawn here, with the associated flow-concentration coupling phenomena discussed by Besseling~\cite{besseling10}.  Concentration effects may possibly be introduced in the cooperativity length or in the yield stress~\cite{Goyon08} of the model, but this would obviously imply the introduction of extra equations accounting for the concentration dynamics.  This surely makes a very interesting  topic for future work.


\section{Conclusions and outlook}\label{sec:conclusions}

It is know that the phenomenology of shear-bandings shows many intriguing effects to be explained~\cite{Coussot02b,Becu,Paredes,Ovarlez08,Moller08,Chikkadi14}. Modeling such a phenomenology is definitively a challenging task~\cite{Mansard11,Martens12,Coussot02c,Coussot10,Martens12,Coussot10,Jagla07,Jagla10,Moorcroft11,Moorcroft13,Varnik1,Varnik2,Chaudhuri,besseling10}. In this paper we have explored the connection between shear-bandings and the cooperativity between local plastic events, characterized by a cooperativity length~\cite{Goyon08,Bocquet09,Goyon10,Geraud13}. This fits in the framework of the fluidity models~\cite{Goyon08,Bocquet09,Goyon10,Geraud13}, which have been proposed in the literature to explain the cooperative flow of complex fluids in confined systems. Technically, based on the simple observation that the fluidity is a non-negative definite order parameter, we have reformulated Bocquet {\it et al.} free-energy functional~\cite{Bocquet09} in terms of its square root $\phi \equiv \pm f^{1/2}$, which is a signed quantity. For the geometry of a Couette flow, it is shown that once the stress-strain constraint $\int_0^1 \phi^2(y) \, dy=S/\sigma$ is taken into account, this functional is minimized by inhomogeneous compact solutions (compactons), which coexist with regions of zero-fluidity (vacuum). Since the latter are unstable, the compactons increase their size at the expense of the vacuum, until they reach the size of the full system, at which point no further decrease of the energy can be achieved, other than by recovering the homogeneous HB solution, namely a linear Couette profile. This leads to a host of remarkable dynamical effects, primarily aging. Given the highly non-trivial dynamics of the fluidity field, we have further investigated how such dynamics is affected by the presence of stochastic fluctuations, typically in the form of mechanical noise.  We wish to emphasize that such noise is inherently non-thermal in character and must be regarded as spontaneous fluidity fluctuations which occur also in the absence of any external load. To be noted that, since the fluidity is non-negative, so must be its average, which implies non-trivial restrictions on the noise strength. Such restrictions do not apply to the square-root fluidity, which, being signed, is free to fluctuate around zero. It is shown that, owing to non-trivial renormalization effects, if the noise is sufficiently strong, the unstable vacuum is stabilized, thereby paving the route to stable shear-band configurations. This qualitative picture is confirmed by numerical simulations of steepest-descent dynamics under different initial conditions and noise amplitudes. The two starting ingredients of the present picture, {\em i.e.,} fluidity as an order parameter and mechanical noise as a promoter of escapes from free-energy minima, are not new. The emerging picture, however, definitely is. In particular, it provides a transparent link between shear-bands and compactons, as well as a subtle stabilization mechanism via non-trivial noise renormalization effects. Finally, we remark that the above rich picture depends only on two free parameters: the cooperative length $\xi$ and the noise strength $\epsilon$. Many further directions for future research can be envisaged. As we have already pointed out earlier on, we have chosen a very simple form of the noise to compute analytically the renormalization effects of the free energy, and we believe the qualitative picture emerging is quite robust and independent (qualitatively) of the structure of the noise. This said, it is of course of decided interest to explore also the dynamic heterogeneity effects brought about by space/time correlations: recent work has highlighted strong correlations in the fluidity fluctuations~\cite{Lamaitre09,SalernoRobbins13,LiuFerreroarXiv2015} and these correlations are likely to exert a significant effect on the flow response near the yielding point~\cite{Budrikis-PRE2013,Lin-PNAS2014,LiuFerreroarXiv2015,Puosi-SoftMatter2015,STZbouchbinder}. It is also of interest to consider the effects of shear-dependent fluidity fluctuations: note in fact that noise can also be triggered by the shear dynamics itself and through the elastic response of the material to local fluidity rearrangements. For example, one could enrich the noise with a dependency on the plastic activity, as proposed in the Hebraud-Lequeux framework~\cite{Hebraud} that underlies the KEP model presented in Bocquet {\it et al.}~\cite{Bocquet09}.\\
Another important issue to be investigated in the present model is the role of boundary conditions. We wish to point out that the boundary conditions may serve as a trigger for the heterogeneity~\cite{Picard01}, hence they have a strong bearing on the onset and stability of compact configurations. In the present work, we have focused on possibly the simplest boundary conditions compatible with compact solutions, although we expect that more general classes of compact fluidity configurations should arise in the presence of different boundary conditions. This is an interesting topic which warrants a separate study on its own for the future.\\

The authors kindly acknowledge funding from the European Research Council under the European Community's Seventh Framework Programme (FP7/2007-2013)/ERC Grant Agreement No. 279004. The authors thank E. Bouchbinder and J.-L. Barrat for useful discussions.

\appendix

\section{Stability of Compactons}\label{appendix1}

To simplify matters in the stability analysis we rescale the space and
order parameter variables as
\begin{equation}
\label{scaling}
y =\frac{\xi \tilde{y}}{\sqrt{m}} \hspace{.2in} \phi=m \tilde{\phi}
\end{equation}
so that the free-energy becomes
\begin{equation}\label{Fa}
F[\tilde{\phi}] = 2 m^{9/2} \xi \int_{y_0}^{y_0+2l_c} \left[-\frac{1}{4} \tilde{\phi}^4
  + \frac{1}{5} \tilde{\phi}^4 |\tilde{\phi}| + \tilde{\phi}^2
  (\partial_{\tilde{y}} \tilde{\phi})^2 \right] d \tilde{y}.
\end{equation}
Then equations \eqref{phi0} and \eqref{E1} become, respectively:
\begin{equation}
\begin{cases}
2 \tilde{\phi}^2 \partial_{\tilde{y}\tilde{y}} \tilde{\phi} + 2 \tilde{\phi} (\partial_{\tilde{y}} \tilde{\phi})^2 + \tilde{\phi}^3 - \tilde{\phi}^3 |\tilde{\phi}| = 0
\\ \tilde{\phi}^2 (\partial_{\tilde{y}} \tilde{\phi})^2 = \tilde{E} -
\frac{1}{4} \tilde{\phi}^4 + \frac{1}{5} \tilde{\phi}^5
\end{cases}
\end{equation}
where the constant $E$ has been rescaled by a factor $m^5$, {\em i.e.,} $\tilde{E}=E/m^5$. To assess the stability of the compactons $\tilde{\phi}_E(\tilde{y})$, we compute $ \delta F \equiv F[\tilde{\phi}_E+\delta \tilde{\phi}]-F[\tilde{\phi}_E]$ up to second order terms in $\delta \tilde{\phi}$. Based on \eqref{phi1}, stability is guaranteed if $\delta F$ is positive defined. After a rather straightforward computation we obtain ($\tilde{\phi}_E>0$)
\begin{equation}
\delta F = 2 m^{9/2} \xi \int_{\tilde{y}_0}^{\tilde{y}_0+2\tilde{l}_c} (\delta \tilde{\phi})^2 \left[ (\partial_{\tilde{y}} \tilde{\phi}_E)^2   -\frac{1}{2} \tilde{\phi}_E^2 + \tilde{\phi}_E^3 \right] \, d\tilde{y} + F_+
\label{s1}
\end{equation}
where $F_+$ includes all positive defined terms
$$
F_+=2 m^{9/2} \xi \int_{\tilde{y}_0}^{\tilde{y}_0+2\tilde{l}_c} \tilde{\phi}^2 (\partial_{\tilde{y}} \delta \tilde{\phi})^2 \, d\tilde{y}.
$$
The integral in \eqref{s1} is computed in the interval $ [\tilde{y}_0, \tilde{y}_0 + 2\tilde{l}_c] $ where the localized solution is defined. Because of symmetry, it is enough to show that the integral between $[\tilde{y}_0, \tilde{y}_0+\tilde{l}_c]$ is positive defined.  Furthermore, we can always choose the origin of integration with the position $\tilde{y}_0$ where the localized solution $\tilde{\phi}_E$ becomes zero and write
$$
\tilde{\phi}_E(\tilde{y}) = (4\tilde{E})^{1/4} A(\tilde{y}) \tilde{y}^{1/2}
$$
where $A$ is an analytic function of $\tilde{y}$. Since $A\partial_{\tilde{y}} A +\tilde{y} (\partial_{\tilde{y}} A )^2 \ge 0$ for $0 \le \tilde{y} \le \tilde{l}_c$ and $F_+$ is positive defined, we obtain
$$
\frac{\delta F}{4 m^{9/2} \xi} \ge \int_0^{\tilde{l}_c} (\delta \tilde{\phi})^2 A^2 \sqrt{4\tilde{E}}\left[\frac{1}{4 \tilde{y}}-\frac{\tilde{y}}{2}+\tilde{y}^{3/2} (4\tilde{E})^{1/4} A({\tilde{y}})\right] \, d\tilde{y}.
$$
We have then studied the properties of $A(\tilde{y})$, finding a lower bound for it: $A(\tilde{y})\ge 1/\sqrt{2}$. Hence
\be
\begin{split}\label{s2}
\frac{\delta F}{4 m^{9/2} \xi} \ge \int_0^{\tilde{l}_c} (\delta \tilde{\phi})^2 A^2\sqrt{4\tilde{E}}\left[\frac{1}{4 \tilde{y}}-\frac{\tilde{y}}{2}+ \frac{\tilde{y}^{3/2}}{\sqrt{2}}(4\tilde{E})^{1/4} \right] \, d\tilde{y} .
\end{split}
\ee
Because the function $[\frac{1}{4\tilde{y}}-\frac{\tilde{y}}{2}+\tilde{y}^{3/2}(4\tilde{E})^{1/4}{1 \over \sqrt{2} } ]$ is positive defined for $\tilde{E}\ge E_c \equiv 0.01$, it follows that all localized solutions with integration constant $\tilde{E}$ in the range $ \tilde{E} \in [0.01, 0.05]$ are local minima of $F[\phi]$, where $\tilde{E}=0.05$ corresponds to the upper bound $E_M=m^5/20$ discussed in Section \ref{sec:model}, above which no compact solution can be found. It is clear from our result that the number of possible stable compactons is still very large.

\section{Self-Consistent Field approximation}\label{appendix2}

In this appendix, we derive the expressions \eqref{eq:scaling0}-\eqref{eq:scaling1} discussed in Section \eqref{sec:model}. We start with equation
\eqref{eq:modelequation}
\begin{equation}
\begin{cases} \partial_t \phi = - \frac{\delta F}{\delta \phi} + \sqrt{\epsilon} \mathrm{w}(y,t) \\
F[\phi]= 2 \int_0^1 \left[-\frac{1}{4} m \phi^4 + \frac{1}{5} \phi^4
  |\phi|+ \xi^2 \phi^2 (\partial_y \phi)^2 \right] d y
\end{cases}
\end{equation}
which we rewrite in the form
\begin{equation} \label{a1}
\begin{cases}
\partial_t \phi = -2 \frac{d V}{d \phi}+4 \xi^2 \phi^2 \partial_{yy} \phi + 4 \xi^2 \phi (\partial_y \phi)^2  + \sqrt{\epsilon} \, \mathrm{w}(y,t) \\ V(\phi) = -\frac{1}{4} m \phi^4 + \frac{1}{5}
\phi^4 |\phi|.
\end{cases}
\end{equation}
Our aim is to show that, because of the noise, $V$ is renormalized and, moreover, the renormalization is state dependent. The difficult terms to be estimated are those proportional to $\xi^2$ in equation \eqref{a1}, {\em i.e.,} $\phi^2 \partial_{yy} \phi$ and $\phi (\partial_y \phi)^2$. Starting from the free-energy, a self-consistent Hartree-like field approximation~\cite{TarziaConiglio1,TarziaConiglio2,Chaikin} can be performed on the term $\phi^2 (\partial_y \phi)^2$
\begin{equation}
\label{a7}
F[\phi] = 2 \int \left[-{1 \over 4} m \phi^4 + {1 \over 5} \phi^4
  |\phi |+ \frac{R}{2}\phi^2+ \frac{D}{2}(\partial_y \phi)^2 \right]\,dy
\end{equation}
where
\begin{equation}
R \equiv 2 \xi^2 \langle (\partial_y \phi )^2 \rangle \hspace{.2in} D
\equiv 2 \xi^2 \langle \phi^2 \rangle.
\end{equation}
Note that formally we should apply the same approximation to the terms $\phi^4$ and $\phi^4 |\phi |$ but for our purpose this would add only small perturbations to the results below described. Next, by using equation \eqref{a7}, we can rewrite equation \eqref{a1} as
\begin{equation}\label{a8}
\partial_t \phi =  -2 \frac{d V_{\mbox{\tiny{eff}}}}{d \phi} +2 D \partial_{yy} \phi+ \sqrt{\epsilon} \, \mathrm{w}(y,t)
\end{equation}
where the {\it effective} potential $V_{\mbox{\tiny{eff}}}$ is given
by
\begin{equation}\label{a9}
V_{\mbox{\tiny{eff}}} (\phi) =  -\frac{1}{4} m \phi^4 + \frac{1}{5} \phi^4 |\phi|+\frac{R}{2} \phi^2.
\end{equation}
According to \eqref{a9}, the effective potential has local minima which can be found by solving the equation ($\phi >0$):
\begin{equation}\label{a12}
\begin{cases}
\left. \frac{d V_{\mbox{\tiny{eff}}}}{d \phi}
\right|_{\phi=\phi_{\mbox{\tiny{min}}}}=\left. \left(R\phi -m \phi^3 +
\phi^4\right) \right|_{\phi=\phi_{\mbox{\tiny{min}}}}=0 \\
\left. \frac{d^2 V_{\mbox{\tiny{eff}}}}{d \phi^2}
\right|_{\phi=\phi_{\mbox{\tiny{min}}}}>0.
\end{cases}
\end{equation}
It is easy to show that \eqref{a12} has one solution $\phi_{\mbox{\tiny{min}}}=\phi_0=0$ for $m \le m_c \sim (27R/4)^{1/3}$, whereas for $m > m_c$ there exist two local minima at $\phi_{\mbox{\tiny{min}}}=\phi_0=0$ and $\phi_{\mbox{\tiny{min}}}=\phi_m \sim m-R/m^2 + O((R/m^2)^2)$ (see top panel in Figure \ref{fig:sketch} for a sketch). To close the problem, we need to compute $R = 2 \xi^2 \langle (\partial_y \phi)^2 \rangle$.  This computation can be done perturbatively ({\em i.e.,} assuming $\epsilon$ to be small) by linearizing equation \eqref{a9} near the minima.\\
We start with the minima at $\phi=\phi_{m} \approx m-R/m^2 + O((R/m^2)^2)$. As we shall see in following, $R$ is proportional to $\epsilon$ and we can simplify the computation by assuming $\phi= m$. Next, we consider the fluctuations $\delta \phi $ near $\phi=m$ which obey the stochastic differential equation
\begin{equation}\label{linear}
\partial_t \delta \phi = 4 \xi^2 m^2 \partial_{yy} \delta \phi - 2 (m^3+R) \delta
\phi + \sqrt{ \epsilon } \, \mathrm{w}(y,t).
\end{equation}
The above linear equation can be solved in Fourier Transform for $\delta \phi_k$ and we obtain
\begin{equation}
\partial_t \delta \phi_k = (-4 \xi^2 m^2 k^2 -2 m^3 - 2R) \delta \phi_k +
\sqrt{\epsilon}\, \mathrm{w}(k,t)
\label{linearnoise}
\end{equation}
where now both $\delta \phi_k \equiv \frac{1}{\sqrt{2 \pi}} \int \exp(iky) \delta \phi(y,t)\,dy$ and $\mathrm{w}(k,t)$ are complex variables and moreover $\langle \mathrm{w}(k_1,t_1) w^*(k_2,t_2) \rangle = \delta(k_1+k_2)\delta(t_1-t_2) $. Equation \eqref{linearnoise} is a linear Langevin equation for each $k$ whose asymptotic variance can be easily computed. After a simple algebra we obtain:
\begin{equation}\label{a12bis}
\begin{split}
R & =  2 \xi^2 \int k^2 \langle \delta \phi_k \delta \phi_k^*\rangle\, dk = \frac{2 \epsilon}{m^2} \int \frac{\xi^2 k^2}{2\xi^2k^2+m+R/m^2} \,
dk \\
& = \frac{\epsilon k_M}{m^2}(1+O(m/k_M)+O( k_M R/m^2)).
\end{split}
\end{equation}
Upon recalling that $m_c^3 \sim R$ and substituting for $R$ the result obtained in \eqref{a12bis}, we obtain $m_c^3 \sim \frac{ \epsilon k_M}{m_c^2}$, and thus \begin{equation}\label{eq:primoscaling} m_c \sim (\epsilon k_M)^{1/5}.
\end{equation}
We refer to this value of $m_c$ as $m^{(1)}$. Our derivation implies that for $m<m^{(1)}$ the only minimum of the effective potential is $\phi_0$ while for $m>m^{(1)}$ there exist two local minima, and in particular the minimum near $\phi=m$ is the global minimum of the potential.\\
Next we consider the minima $\phi=0$. In this case the problem is more complicated since $R$ should be computed self-consistently with $D$. To solve the problem, we linearize equation \eqref{a9} near $\phi=0$ obtaining:
\begin{equation}\label{a13}
\partial_t \delta \phi = 2D \partial_{yy} \delta \phi - 2R \delta \phi +
\sqrt{\epsilon}\, \mathrm{w}(y,t).
\end{equation}
We next perform a Fourier transform obtaining a set of linear Langevin equations. A simple algebra gives:
\begin{equation}\label{DandR}
D = 2 \xi^2 \int \langle \delta \phi_k \delta \phi_k^*\rangle \, dk =
2 \epsilon \xi^2 \int \frac{dk}{Dk^2+R}
\end{equation}
\begin{equation}\label{a14}
R = 2 \xi^2 \int k^2 \langle \delta \phi_k \delta \phi_k^*\rangle \,dk
= 2 \epsilon \xi^2 \int \frac{ k^2 dk}{Dk^2+R}.
\end{equation}
By multiplying equation \eqref{DandR} by $R$ and equation \eqref{a14} by $D$, and summing the two results, we get
\begin{equation}\label{eq:cons1}
2DR=2\epsilon \xi^2 k_M.
\end{equation}
Another relation between $D$ and $R$ can be explicitly obtained by solving exactly the integral in \eqref{DandR} and assuming $k_M$ large
\begin{equation}\label{eq:cons2}
\begin{split}
D & = 2\epsilon \xi^2 \int \frac{dk}{Dk^2+R}= 2\frac{\epsilon \xi^2}{R} \int \frac{dk}{Dk^2/R+1}\\
& =2\frac{\epsilon \xi^2}{R} \sqrt{\frac{R}{D}} \mbox{atan} \left(\sqrt{\frac{D}{R}} k_M \right) \approx \frac{\pi \epsilon \xi^2}{R} \sqrt{\frac{R}{D}}.
\end{split}
\end{equation}
The self-consistent solutions of \eqref{eq:cons1}-\eqref{eq:cons2} are given by:
\begin{equation}
\label{a15}
D \sim \frac{\xi \sqrt{\epsilon k_M}} {k_M} ; \,\,\,\, R \sim \xi k_M
\sqrt{\epsilon k_M}.
\end{equation}
Equation \eqref{a15} implies that the fluctuations near $\phi_{0}$ are given by
\begin{equation}\label{a16pre}
\langle (\delta \phi)^2 \rangle = \frac{D}{\xi^2} \sim \frac{\sqrt{\epsilon k_M}}{\xi k_M}
\end{equation}
meaning that near $\phi_{0}$ the average fluidity $f_0$ is not zero, which is the result reported in \eqref{f0}. Finally, we recall that $m_c ^3 \sim R$, so that
\begin{equation}\label{a16}
m_c \sim (\xi k_M \sqrt{\epsilon k_M})^{1/3}
\end{equation}
and we refer to this value of $m_c$ as $m^{(0)}$.\\

In order to gain a physical insight in the above approximations, it is expedient to consider again equation \eqref{a1} and decompose $\phi$ as follows:
\begin{equation}
\phi=\langle \phi \rangle + \delta \phi
\end{equation}
where $\delta \phi$ are (small) fluctuations around the space average $\langle \phi \rangle$. Equation \eqref{a1} is then rewritten as:
\begin{equation}
\partial_t \phi = 2 m \phi^3 - 2 \phi^3 |\phi| +  4 \xi^2 \partial_y [ \phi^2( \partial_y \phi)] - 4 \xi^2 \phi (\partial_y \phi)^2 + \sqrt{\epsilon}
\mathrm{w}(y,t)
\label{b1}
\end{equation}
Upon performing the space average of \eqref{b1} and assuming self-averaging ({\em i.e.,} ensemble average is equal to space average), we obtain:
\begin{equation}\label{b2}
\partial_t \langle \phi \rangle =  2 m \langle \phi \rangle^3 - 2 \langle \phi \rangle^3 |\langle \phi \rangle|- 4 \xi^2 \langle \phi \rangle \langle (\partial_y \delta \phi)^2 \rangle
\end{equation}
where we have neglected terms proportional to $(\delta\phi) (\partial_y \delta \phi)^2 $ and also terms coming from the decomposition of $\phi^3$ and $\phi^3 | \phi |$, which are supposedly much smaller than $(\partial_y \delta \phi)^2$. By linearizing the rhs of \eqref{b1} around $\langle \phi \rangle > 0$, and subtracting equation \eqref{b2}, we obtain (to the linear order in $\delta \phi$) the following equation
\begin{equation}\label{b3}
\partial_t \delta \phi = 2 (3 m \langle \phi \rangle^2 - 4 \langle \phi \rangle^3) \delta \phi + 4 \xi^2 \langle \phi \rangle^2 \partial_{yy} \delta \phi + \sqrt{\epsilon}\, \mathrm{w}(y,t).
\end{equation}
Next, assuming that $\langle \phi \rangle \sim m + O(\epsilon)$, we can compute the quantity $R = 2 \xi^2 \langle (\partial_y \delta \phi)^2 \rangle $ from (\ref{b3}): the equation is linear and can be solved similarly to what we have done for \eqref{linear}, obtaining the ${\cal O}(\epsilon)$ result in \eqref{a12bis}.\\ Moreover, we can recover the results for $m^{(0)}$ by writing equation \eqref{a1} for the fluidity $f= \phi^2$. Upon using Ito calculus~\cite{Risken}, we obtain:
\begin{equation}
\partial_t f = \epsilon k_M + O( f^2) + \xi^2 f \partial_{yy} f + \sqrt{\epsilon f}\, \mathrm{w}(y,t)
\label{b4}
\end{equation}
where the terms $O(f^2)$ are neglected in the following since we are investigating the stability of the state $\phi=0$. Upon averaging in space (\ref{b4}) and assuming self-averaging to hold, we find
\begin{equation}
\partial_t \langle f \rangle = - \xi^2 \langle (\partial_y f )^2 \rangle+ \epsilon k_M.
\label{b5}
\end{equation}
We then use the approximation $\langle (\partial_y f)^2 \rangle = \alpha k^2_M \langle f \rangle^2 $, where $\alpha$ is a constant of order $1$. At the stationary state, equation \eqref{b5} predicts
\begin{equation}
\xi^2 \alpha  k^2_M \langle f \rangle ^2 = \epsilon k_M
\end{equation}
which is equivalent to
\begin{equation}\label{b6}
\langle (\delta \phi)^2 \rangle = \frac{\sqrt{\epsilon k_M}}{\alpha^{1/2}\xi k_M}.
\end{equation}
Equation \eqref{b6}, apart for a numerical constant, gives exactly the same result reported in \eqref{a15}-\eqref{a16pre}.

\bibliography{rsc}       

\providecommand*{\mcitethebibliography}{\thebibliography}
\csname @ifundefined\endcsname{endmcitethebibliography}
{\let\endmcitethebibliography\endthebibliography}{}
\begin{mcitethebibliography}{83}
\providecommand*{\natexlab}[1]{#1}
\providecommand*{\mciteSetBstSublistMode}[1]{}
\providecommand*{\mciteSetBstMaxWidthForm}[2]{}
\providecommand*{\mciteBstWouldAddEndPuncttrue}
  {\def\EndOfBibitem{\unskip.}}
\providecommand*{\mciteBstWouldAddEndPunctfalse}
  {\let\EndOfBibitem\relax}
\providecommand*{\mciteSetBstMidEndSepPunct}[3]{}
\providecommand*{\mciteSetBstSublistLabelBeginEnd}[3]{}
\providecommand*{\EndOfBibitem}{}
\mciteSetBstSublistMode{f}
\mciteSetBstMaxWidthForm{subitem}
{(\emph{\alph{mcitesubitemcount}})}
\mciteSetBstSublistLabelBeginEnd{\mcitemaxwidthsubitemform\space}
{\relax}{\relax}

\bibitem[Larson(1999)]{Larson}
R.~G. Larson, \emph{The Structure and Rheology of Complex Fluids}, Oxford
  University Press, 1999\relax
\mciteBstWouldAddEndPuncttrue
\mciteSetBstMidEndSepPunct{\mcitedefaultmidpunct}
{\mcitedefaultendpunct}{\mcitedefaultseppunct}\relax
\EndOfBibitem
\bibitem[Voigtmann(2014)]{Voigtmann}
T.~Voigtmann, \emph{Current Opinion in Colloid \& Interface Science}, 2014,
  \textbf{19}, 549--560\relax
\mciteBstWouldAddEndPuncttrue
\mciteSetBstMidEndSepPunct{\mcitedefaultmidpunct}
{\mcitedefaultendpunct}{\mcitedefaultseppunct}\relax
\EndOfBibitem
\bibitem[Coussot(2005)]{Coussot}
P.~Coussot, \emph{Rheometry of Pastes, Suspensions, and Granular Materials},
  Wiley-Interscience, 2005\relax
\mciteBstWouldAddEndPuncttrue
\mciteSetBstMidEndSepPunct{\mcitedefaultmidpunct}
{\mcitedefaultendpunct}{\mcitedefaultseppunct}\relax
\EndOfBibitem
\bibitem[Ragouilliaux \emph{et~al.}(2007)Ragouilliaux, Ovarlez,
  Shahidzadeh-Bonn, Herzhaft, Palermo, and Coussot]{Ragouilliaux07}
A.~Ragouilliaux, G.~Ovarlez, N.~Shahidzadeh-Bonn, B.~Herzhaft, T.~Palermo and
  P.~Coussot, \emph{Phys. Rev. E}, 2007,  051408\relax
\mciteBstWouldAddEndPuncttrue
\mciteSetBstMidEndSepPunct{\mcitedefaultmidpunct}
{\mcitedefaultendpunct}{\mcitedefaultseppunct}\relax
\EndOfBibitem
\bibitem[Divoux \emph{et~al.}(2011)Divoux, Barentin, and
  Manneville]{Manneville2011a}
T.~Divoux, C.~Barentin and S.~Manneville, \emph{Soft Matter}, 2011, \textbf{7},
  9335\relax
\mciteBstWouldAddEndPuncttrue
\mciteSetBstMidEndSepPunct{\mcitedefaultmidpunct}
{\mcitedefaultendpunct}{\mcitedefaultseppunct}\relax
\EndOfBibitem
\bibitem[Ovarlez \emph{et~al.}(2010)Ovarlez, Krishanand, and
  Cohen-Addad]{Ovarlez10}
G.~Ovarlez, K.~Krishanand and S.~Cohen-Addad, \emph{Europhys. Lett.}, 2010,
  \textbf{91}, 68005\relax
\mciteBstWouldAddEndPuncttrue
\mciteSetBstMidEndSepPunct{\mcitedefaultmidpunct}
{\mcitedefaultendpunct}{\mcitedefaultseppunct}\relax
\EndOfBibitem
\bibitem[Becu \emph{et~al.}(2006)Becu, Manneville, and Colin]{Becu}
L.~Becu, S.~Manneville and A.~Colin, \emph{Phys. Rev. Lett.}, 2006,
  \textbf{96}, 138302\relax
\mciteBstWouldAddEndPuncttrue
\mciteSetBstMidEndSepPunct{\mcitedefaultmidpunct}
{\mcitedefaultendpunct}{\mcitedefaultseppunct}\relax
\EndOfBibitem
\bibitem[Princen and Kiss(1989)]{Princen89}
H.~Princen and A.~Kiss, \emph{J. Colloid Interface Sci.}, 1989, \textbf{128},
  176--187\relax
\mciteBstWouldAddEndPuncttrue
\mciteSetBstMidEndSepPunct{\mcitedefaultmidpunct}
{\mcitedefaultendpunct}{\mcitedefaultseppunct}\relax
\EndOfBibitem
\bibitem[Marze \emph{et~al.}(2008)Marze, Langevin, and Saint-Jalmes]{Marze08}
S.~Marze, D.~Langevin and A.~Saint-Jalmes, \emph{J. Rheol.}, 2008, \textbf{52},
  1091--1111\relax
\mciteBstWouldAddEndPuncttrue
\mciteSetBstMidEndSepPunct{\mcitedefaultmidpunct}
{\mcitedefaultendpunct}{\mcitedefaultseppunct}\relax
\EndOfBibitem
\bibitem[Denkov \emph{et~al.}(2009)Denkov, Tcholakova, Golemanov,
  Ananthapadmanabhan, and Lips]{Denkov09}
N.~Denkov, S.~Tcholakova, K.~Golemanov, K.~Ananthapadmanabhan and A.~Lips,
  \emph{Soft Matter}, 2009, \textbf{5}, 3389--3408\relax
\mciteBstWouldAddEndPuncttrue
\mciteSetBstMidEndSepPunct{\mcitedefaultmidpunct}
{\mcitedefaultendpunct}{\mcitedefaultseppunct}\relax
\EndOfBibitem
\bibitem[Coussot \emph{et~al.}(2002)Coussot, Nguyen, Huynh, and
  Bonn]{Coussot02b}
P.~Coussot, Q.~Nguyen, H.~T. Huynh and D.~Bonn, \emph{Phys. Rev. Lett.}, 2002,
  \textbf{88}, 175501\relax
\mciteBstWouldAddEndPuncttrue
\mciteSetBstMidEndSepPunct{\mcitedefaultmidpunct}
{\mcitedefaultendpunct}{\mcitedefaultseppunct}\relax
\EndOfBibitem
\bibitem[Paredes \emph{et~al.}(2011)Paredes, Shahidzadeh-Bonn, and
  Bonn]{Paredes}
J.~Paredes, N.~Shahidzadeh-Bonn and D.~Bonn, \emph{Phys. Rev. Lett.}, 2011,
  \textbf{23}, 28411\relax
\mciteBstWouldAddEndPuncttrue
\mciteSetBstMidEndSepPunct{\mcitedefaultmidpunct}
{\mcitedefaultendpunct}{\mcitedefaultseppunct}\relax
\EndOfBibitem
\bibitem[Ovarlez \emph{et~al.}(2008)Ovarlez, Rodts, Ragouilliaux, Coussot,
  Goyon, and Colin]{Ovarlez08}
G.~Ovarlez, S.~Rodts, A.~Ragouilliaux, P.~Coussot, J.~Goyon and A.~Colin,
  \emph{Phys. Rev. E}, 2008, \textbf{78}, 036307\relax
\mciteBstWouldAddEndPuncttrue
\mciteSetBstMidEndSepPunct{\mcitedefaultmidpunct}
{\mcitedefaultendpunct}{\mcitedefaultseppunct}\relax
\EndOfBibitem
\bibitem[Moller \emph{et~al.}(2008)Moller, Rodts, Michels, and Bonn]{Moller08}
P.~Moller, S.~Rodts, M.~Michels and D.~Bonn, \emph{Phys. Rev. E}, 2008,
  \textbf{77}, 041507\relax
\mciteBstWouldAddEndPuncttrue
\mciteSetBstMidEndSepPunct{\mcitedefaultmidpunct}
{\mcitedefaultendpunct}{\mcitedefaultseppunct}\relax
\EndOfBibitem
\bibitem[Cruz \emph{et~al.}(2002)Cruz, Chevoir, Bonn, and Coussot]{BonnCoussot}
F.~D. Cruz, F.~Chevoir, D.~Bonn and P.~Coussot, \emph{Phys. Rev. E}, 2002,
  \textbf{66}, 051305\relax
\mciteBstWouldAddEndPuncttrue
\mciteSetBstMidEndSepPunct{\mcitedefaultmidpunct}
{\mcitedefaultendpunct}{\mcitedefaultseppunct}\relax
\EndOfBibitem
\bibitem[Sollich \emph{et~al.}(1997)Sollich, Lequeux, Hebraud, and
  Cates]{Sollich1}
P.~Sollich, F.~Lequeux, P.~Hebraud and M.~E. Cates, \emph{Phys. Rev. Lett.},
  1997, \textbf{78}, 2020\relax
\mciteBstWouldAddEndPuncttrue
\mciteSetBstMidEndSepPunct{\mcitedefaultmidpunct}
{\mcitedefaultendpunct}{\mcitedefaultseppunct}\relax
\EndOfBibitem
\bibitem[Sollich(1998)]{Sollich2}
P.~Sollich, \emph{Phys. Rev. E}, 1998, \textbf{58}, 738\relax
\mciteBstWouldAddEndPuncttrue
\mciteSetBstMidEndSepPunct{\mcitedefaultmidpunct}
{\mcitedefaultendpunct}{\mcitedefaultseppunct}\relax
\EndOfBibitem
\bibitem[Sollich and Cates(2012)]{Sollich3}
P.~Sollich and M.~E. Cates, \emph{Phys. Rev. E}, 2012, \textbf{85},
  031127\relax
\mciteBstWouldAddEndPuncttrue
\mciteSetBstMidEndSepPunct{\mcitedefaultmidpunct}
{\mcitedefaultendpunct}{\mcitedefaultseppunct}\relax
\EndOfBibitem
\bibitem[Fielding \emph{et~al.}(2000)Fielding, Sollich, and Cates]{Fielding1}
S.~M. Fielding, P.~Sollich and M.~E. Cates, \emph{J. Rheol.}, 2000,
  \textbf{44}, 323\relax
\mciteBstWouldAddEndPuncttrue
\mciteSetBstMidEndSepPunct{\mcitedefaultmidpunct}
{\mcitedefaultendpunct}{\mcitedefaultseppunct}\relax
\EndOfBibitem
\bibitem[Fielding \emph{et~al.}(2009)Fielding, Cates, and Sollich]{Fielding2}
S.~M. Fielding, M.~E. Cates and P.~Sollich, \emph{Soft Matter}, 2009,
  \textbf{5}, 2378\relax
\mciteBstWouldAddEndPuncttrue
\mciteSetBstMidEndSepPunct{\mcitedefaultmidpunct}
{\mcitedefaultendpunct}{\mcitedefaultseppunct}\relax
\EndOfBibitem
\bibitem[Langer(2008)]{Langer}
J.~S. Langer, \emph{Phys. Rev. E}, 2008, \textbf{77}, 021502\relax
\mciteBstWouldAddEndPuncttrue
\mciteSetBstMidEndSepPunct{\mcitedefaultmidpunct}
{\mcitedefaultendpunct}{\mcitedefaultseppunct}\relax
\EndOfBibitem
\bibitem[Pouliquen and Forterre(2009)]{Pouliquen}
O.~Pouliquen and Y.~Forterre, \emph{Phil. Trans. R. Soc. London A}, 2009,
  \textbf{367}, 5091\relax
\mciteBstWouldAddEndPuncttrue
\mciteSetBstMidEndSepPunct{\mcitedefaultmidpunct}
{\mcitedefaultendpunct}{\mcitedefaultseppunct}\relax
\EndOfBibitem
\bibitem[Derec \emph{et~al.}(2001)Derec, Ajdari, and Lequeux]{Derec}
C.~Derec, A.~Ajdari and F.~Lequeux, \emph{Eur. Phys. J. E}, 2001, \textbf{4},
  355--361\relax
\mciteBstWouldAddEndPuncttrue
\mciteSetBstMidEndSepPunct{\mcitedefaultmidpunct}
{\mcitedefaultendpunct}{\mcitedefaultseppunct}\relax
\EndOfBibitem
\bibitem[Picard \emph{et~al.}(2002)Picard, Ajdari, Bocquet, and
  Lequeux]{Picard01}
G.~Picard, A.~Ajdari, L.~Bocquet and F.~Lequeux, \emph{Phys. Rev. E}, 2002,
  \textbf{66}, 051501\relax
\mciteBstWouldAddEndPuncttrue
\mciteSetBstMidEndSepPunct{\mcitedefaultmidpunct}
{\mcitedefaultendpunct}{\mcitedefaultseppunct}\relax
\EndOfBibitem
\bibitem[Picard \emph{et~al.}(2004)Picard, Ajdari, Lequeux, and
  Bocquet]{Picard}
G.~Picard, A.~Ajdari, F.~Lequeux and L.~Bocquet, \emph{Eur. Phys. J. E}, 2004,
  \textbf{15}, 371--381\relax
\mciteBstWouldAddEndPuncttrue
\mciteSetBstMidEndSepPunct{\mcitedefaultmidpunct}
{\mcitedefaultendpunct}{\mcitedefaultseppunct}\relax
\EndOfBibitem
\bibitem[Mansard \emph{et~al.}(2011)Mansard, Colin, Chauduri, and
  Bocquet]{Mansard11}
V.~Mansard, A.~Colin, P.~Chauduri and L.~Bocquet, \emph{Soft matter}, 2011,
  \textbf{7}, 5524\relax
\mciteBstWouldAddEndPuncttrue
\mciteSetBstMidEndSepPunct{\mcitedefaultmidpunct}
{\mcitedefaultendpunct}{\mcitedefaultseppunct}\relax
\EndOfBibitem
\bibitem[Nicolas and Barrat(2013)]{Nicolas13}
A.~Nicolas and J.-L. Barrat, \emph{Phys. Rev. Lett.}, 2013, \textbf{110},
  138304\relax
\mciteBstWouldAddEndPuncttrue
\mciteSetBstMidEndSepPunct{\mcitedefaultmidpunct}
{\mcitedefaultendpunct}{\mcitedefaultseppunct}\relax
\EndOfBibitem
\bibitem[Mansard \emph{et~al.}(2013)Mansard, Colin, Chauduri, and
  Bocquet]{Mansard13}
V.~Mansard, A.~Colin, P.~Chauduri and L.~Bocquet, \emph{Soft matter}, 2013,
  \textbf{9}, 7489--7500\relax
\mciteBstWouldAddEndPuncttrue
\mciteSetBstMidEndSepPunct{\mcitedefaultmidpunct}
{\mcitedefaultendpunct}{\mcitedefaultseppunct}\relax
\EndOfBibitem
\bibitem[Goyon \emph{et~al.}(2008)Goyon, Colin, Ovarlez, Ajdari, and
  Bocquet]{Goyon08}
J.~Goyon, A.~Colin, G.~Ovarlez, A.~Ajdari and L.~Bocquet, \emph{Nature}, 2008,
  \textbf{454}, 84--87\relax
\mciteBstWouldAddEndPuncttrue
\mciteSetBstMidEndSepPunct{\mcitedefaultmidpunct}
{\mcitedefaultendpunct}{\mcitedefaultseppunct}\relax
\EndOfBibitem
\bibitem[Bocquet \emph{et~al.}(2009)Bocquet, Colin, and Ajdari]{Bocquet09}
L.~Bocquet, A.~Colin and A.~Ajdari, \emph{Phys. Rev. Lett.}, 2009,
  \textbf{103}, 036001\relax
\mciteBstWouldAddEndPuncttrue
\mciteSetBstMidEndSepPunct{\mcitedefaultmidpunct}
{\mcitedefaultendpunct}{\mcitedefaultseppunct}\relax
\EndOfBibitem
\bibitem[Goyon \emph{et~al.}(2010)Goyon, Colin, and Bocquet]{Goyon10}
J.~Goyon, A.~Colin and L.~Bocquet, \emph{Soft Matter}, 2010, \textbf{6},
  2668--2678\relax
\mciteBstWouldAddEndPuncttrue
\mciteSetBstMidEndSepPunct{\mcitedefaultmidpunct}
{\mcitedefaultendpunct}{\mcitedefaultseppunct}\relax
\EndOfBibitem
\bibitem[Geraud \emph{et~al.}(2013)Geraud, Bocquet, and Barentin]{Geraud13}
B.~Geraud, L.~Bocquet and C.~Barentin, \emph{Eur. Phys. J. E}, 2013,
  \textbf{36}, 30\relax
\mciteBstWouldAddEndPuncttrue
\mciteSetBstMidEndSepPunct{\mcitedefaultmidpunct}
{\mcitedefaultendpunct}{\mcitedefaultseppunct}\relax
\EndOfBibitem
\bibitem[Kamrin and Koval(2012)]{Kamrin12}
K.~Kamrin and G.~Koval, \emph{Phys. Rev. Lett.}, 2012, \textbf{108},
  178301\relax
\mciteBstWouldAddEndPuncttrue
\mciteSetBstMidEndSepPunct{\mcitedefaultmidpunct}
{\mcitedefaultendpunct}{\mcitedefaultseppunct}\relax
\EndOfBibitem
\bibitem[Amon \emph{et~al.}(2012)Amon, Nguyen, Bruand, Crassous, and
  Cl\'ement]{Amon12}
A.~Amon, V.~B. Nguyen, A.~Bruand, J.~Crassous and E.~Cl\'ement, \emph{Phys.
  Rev. Lett.}, 2012, \textbf{108}, 135502\relax
\mciteBstWouldAddEndPuncttrue
\mciteSetBstMidEndSepPunct{\mcitedefaultmidpunct}
{\mcitedefaultendpunct}{\mcitedefaultseppunct}\relax
\EndOfBibitem
\bibitem[Katgert \emph{et~al.}(2010)Katgert, Tighe, Mobius, and
  Hecke]{Katgert10}
G.~Katgert, B.~Tighe, M.~Mobius and M.~V. Hecke, \emph{Europhys. Lett.}, 2010,
  \textbf{90}, 54002\relax
\mciteBstWouldAddEndPuncttrue
\mciteSetBstMidEndSepPunct{\mcitedefaultmidpunct}
{\mcitedefaultendpunct}{\mcitedefaultseppunct}\relax
\EndOfBibitem
\bibitem[Sbragaglia \emph{et~al.}(2012)Sbragaglia, Benzi, Bernaschi, and
  Succi]{Sbragaglia12}
M.~Sbragaglia, R.~Benzi, M.~Bernaschi and S.~Succi, \emph{Soft Matter}, 2012,
  \textbf{8}, 10773--10782\relax
\mciteBstWouldAddEndPuncttrue
\mciteSetBstMidEndSepPunct{\mcitedefaultmidpunct}
{\mcitedefaultendpunct}{\mcitedefaultseppunct}\relax
\EndOfBibitem
\bibitem[Dollet \emph{et~al.}(2015)Dollet, Scagliarini, and Sbragaglia]{JFM}
B.~Dollet, A.~Scagliarini and M.~Sbragaglia, \emph{Journal of Fluid Mechanics},
  2015, \textbf{766}, 556--589\relax
\mciteBstWouldAddEndPuncttrue
\mciteSetBstMidEndSepPunct{\mcitedefaultmidpunct}
{\mcitedefaultendpunct}{\mcitedefaultseppunct}\relax
\EndOfBibitem
\bibitem[Mansard \emph{et~al.}(2014)Mansard, Colin, and Bocquet]{Mansard14}
V.~Mansard, A.~Colin and L.~Bocquet, \emph{Soft matter}, 2014, \textbf{10},
  6984--6989\relax
\mciteBstWouldAddEndPuncttrue
\mciteSetBstMidEndSepPunct{\mcitedefaultmidpunct}
{\mcitedefaultendpunct}{\mcitedefaultseppunct}\relax
\EndOfBibitem
\bibitem[Moorcroft \emph{et~al.}(2011)Moorcroft, Cates, and
  Fielding]{Fielding3}
R.~L. Moorcroft, M.~E. Cates and S.~M. Fielding, \emph{Phys. Rev. Lett.}, 2011,
  \textbf{106}, 055502\relax
\mciteBstWouldAddEndPuncttrue
\mciteSetBstMidEndSepPunct{\mcitedefaultmidpunct}
{\mcitedefaultendpunct}{\mcitedefaultseppunct}\relax
\EndOfBibitem
\bibitem[Varnik \emph{et~al.}(2004)Varnik, Bocquet, and Barrat]{Varnik1}
F.~Varnik, L.~Bocquet and J.-L. Barrat, \emph{Jour. Chem. Phys.}, 2004,
  \textbf{120}, 2788\relax
\mciteBstWouldAddEndPuncttrue
\mciteSetBstMidEndSepPunct{\mcitedefaultmidpunct}
{\mcitedefaultendpunct}{\mcitedefaultseppunct}\relax
\EndOfBibitem
\bibitem[Varnik \emph{et~al.}(2003)Varnik, Bocquet, Barrat, and
  Berthier]{Varnik2}
F.~Varnik, L.~Bocquet, J.-L. Barrat and L.~Berthier, \emph{Phys. Rev. Lett.},
  2003, \textbf{90}, 095702\relax
\mciteBstWouldAddEndPuncttrue
\mciteSetBstMidEndSepPunct{\mcitedefaultmidpunct}
{\mcitedefaultendpunct}{\mcitedefaultseppunct}\relax
\EndOfBibitem
\bibitem[Fielding(2014)]{Fielding14}
S.~M. Fielding, \emph{Rep. Prog. Phys.}, 2014, \textbf{77}, 102601\relax
\mciteBstWouldAddEndPuncttrue
\mciteSetBstMidEndSepPunct{\mcitedefaultmidpunct}
{\mcitedefaultendpunct}{\mcitedefaultseppunct}\relax
\EndOfBibitem
\bibitem[Bouchbinder and Langer(2009)]{Bouchbinder1}
E.~Bouchbinder and J.~S. Langer, \emph{Phys. Rev. E}, 2009, \textbf{80},
  031131\relax
\mciteBstWouldAddEndPuncttrue
\mciteSetBstMidEndSepPunct{\mcitedefaultmidpunct}
{\mcitedefaultendpunct}{\mcitedefaultseppunct}\relax
\EndOfBibitem
\bibitem[Bouchbinder and Langer(2009)]{Bouchbinder2}
E.~Bouchbinder and J.~S. Langer, \emph{Phys. Rev. E}, 2009, \textbf{80},
  031132\relax
\mciteBstWouldAddEndPuncttrue
\mciteSetBstMidEndSepPunct{\mcitedefaultmidpunct}
{\mcitedefaultendpunct}{\mcitedefaultseppunct}\relax
\EndOfBibitem
\bibitem[Bouchbinder and Langer(2009)]{Bouchbinder3}
E.~Bouchbinder and J.~S. Langer, \emph{Phys. Rev. E}, 2009, \textbf{80},
  031133\relax
\mciteBstWouldAddEndPuncttrue
\mciteSetBstMidEndSepPunct{\mcitedefaultmidpunct}
{\mcitedefaultendpunct}{\mcitedefaultseppunct}\relax
\EndOfBibitem
\bibitem[Gittings and Durian(2008)]{Durian08}
A.~S. Gittings and D.~J. Durian, \emph{Phys. Rev. E}, 2008, \textbf{78},
  066313\relax
\mciteBstWouldAddEndPuncttrue
\mciteSetBstMidEndSepPunct{\mcitedefaultmidpunct}
{\mcitedefaultendpunct}{\mcitedefaultseppunct}\relax
\EndOfBibitem
\bibitem[Roth \emph{et~al.}(2013)Roth, Jones, and Durian]{Durian13}
A.~E. Roth, C.~D. Jones and D.~J. Durian, \emph{Phys. Rev. E}, 2013,
  \textbf{87}, 042304\relax
\mciteBstWouldAddEndPuncttrue
\mciteSetBstMidEndSepPunct{\mcitedefaultmidpunct}
{\mcitedefaultendpunct}{\mcitedefaultseppunct}\relax
\EndOfBibitem
\bibitem[Benzi \emph{et~al.}(2015)Benzi, Sbragaglia, Perlekar, Bernaschi,
  Succi, and Toschi]{SoftMatter15}
R.~Benzi, M.~Sbragaglia, P.~Perlekar, M.~Bernaschi, S.~Succi and F.~Toschi,
  \emph{Soft Matter}, 2015, \textbf{11}, 1271\relax
\mciteBstWouldAddEndPuncttrue
\mciteSetBstMidEndSepPunct{\mcitedefaultmidpunct}
{\mcitedefaultendpunct}{\mcitedefaultseppunct}\relax
\EndOfBibitem
\bibitem[Tarzia and Coniglio(2006)]{TarziaConiglio1}
M.~Tarzia and A.~Coniglio, \emph{Phys. Rev. Lett.}, 2006, \textbf{96},
  075702\relax
\mciteBstWouldAddEndPuncttrue
\mciteSetBstMidEndSepPunct{\mcitedefaultmidpunct}
{\mcitedefaultendpunct}{\mcitedefaultseppunct}\relax
\EndOfBibitem
\bibitem[Tarzia and Coniglio(2007)]{TarziaConiglio2}
M.~Tarzia and A.~Coniglio, \emph{Phys. Rev. E}, 2007, \textbf{75}, 011410\relax
\mciteBstWouldAddEndPuncttrue
\mciteSetBstMidEndSepPunct{\mcitedefaultmidpunct}
{\mcitedefaultendpunct}{\mcitedefaultseppunct}\relax
\EndOfBibitem
\bibitem[Chaikin and Lubensky(1995)]{Chaikin}
P.~M. Chaikin and T.~C. Lubensky, \emph{Principles of Condensed Matter
  Physics}, Cambridge University Press, 1995\relax
\mciteBstWouldAddEndPuncttrue
\mciteSetBstMidEndSepPunct{\mcitedefaultmidpunct}
{\mcitedefaultendpunct}{\mcitedefaultseppunct}\relax
\EndOfBibitem
\bibitem[Boukany and Wang(2010)]{Boukany10}
P.~Boukany and S.-Q. Wang, \emph{Macromolecules}, 2010, \textbf{43}, 6950\relax
\mciteBstWouldAddEndPuncttrue
\mciteSetBstMidEndSepPunct{\mcitedefaultmidpunct}
{\mcitedefaultendpunct}{\mcitedefaultseppunct}\relax
\EndOfBibitem
\bibitem[Ravindranath \emph{et~al.}(2008)Ravindranath, Wang, Ofechnowicz, and
  Quirk]{Ravindranath08}
S.~Ravindranath, S.-Q. Wang, M.~Ofechnowicz and R.~Quirk,
  \emph{Macromolecules}, 2008, \textbf{41}, 2663\relax
\mciteBstWouldAddEndPuncttrue
\mciteSetBstMidEndSepPunct{\mcitedefaultmidpunct}
{\mcitedefaultendpunct}{\mcitedefaultseppunct}\relax
\EndOfBibitem
\bibitem[Tapadia and Wang(2006)]{Tapadia06}
P.~Tapadia and S.~Wang, \emph{Phys. Rev. Lett.}, 2006, \textbf{96},
  016001\relax
\mciteBstWouldAddEndPuncttrue
\mciteSetBstMidEndSepPunct{\mcitedefaultmidpunct}
{\mcitedefaultendpunct}{\mcitedefaultseppunct}\relax
\EndOfBibitem
\bibitem[Boukany and Wang(2008)]{Boukany08}
P.~E. Boukany and S.-Q. Wang, \emph{Macromolecules}, 2008, \textbf{41},
  1455\relax
\mciteBstWouldAddEndPuncttrue
\mciteSetBstMidEndSepPunct{\mcitedefaultmidpunct}
{\mcitedefaultendpunct}{\mcitedefaultseppunct}\relax
\EndOfBibitem
\bibitem[Britton and Callaghan(1997)]{Britton97}
M.~M. Britton and P.~T. Callaghan, \emph{Phys. Rev. Lett.}, 1997, \textbf{78},
  4930\relax
\mciteBstWouldAddEndPuncttrue
\mciteSetBstMidEndSepPunct{\mcitedefaultmidpunct}
{\mcitedefaultendpunct}{\mcitedefaultseppunct}\relax
\EndOfBibitem
\bibitem[Helgeson \emph{et~al.}(2009)Helgeson, Reichert, Hu, and
  Wagner]{Helgeson09}
M.~E. Helgeson, M.~D. Reichert, Y.~T. Hu and N.~J. Wagner, \emph{Soft Matter},
  2009, \textbf{5}, 3858\relax
\mciteBstWouldAddEndPuncttrue
\mciteSetBstMidEndSepPunct{\mcitedefaultmidpunct}
{\mcitedefaultendpunct}{\mcitedefaultseppunct}\relax
\EndOfBibitem
\bibitem[Salmon \emph{et~al.}(2003)Salmon, Colin, Manneville, and
  Molino]{Salmon03}
J.~Salmon, A.~Colin, S.~Manneville and F.~Molino, \emph{Phys. Rev. Lett.},
  2003, \textbf{90}, 228303\relax
\mciteBstWouldAddEndPuncttrue
\mciteSetBstMidEndSepPunct{\mcitedefaultmidpunct}
{\mcitedefaultendpunct}{\mcitedefaultseppunct}\relax
\EndOfBibitem
\bibitem[Manning \emph{et~al.}(2009)Manning, Daub, Langer, and Carlson]{STZ}
M.~Manning, E.~G. Daub, J.~S. Langer and J.~M. Carlson, \emph{Phys. Rev. E},
  2009, \textbf{79}, 016110\relax
\mciteBstWouldAddEndPuncttrue
\mciteSetBstMidEndSepPunct{\mcitedefaultmidpunct}
{\mcitedefaultendpunct}{\mcitedefaultseppunct}\relax
\EndOfBibitem
\bibitem[Dhont and Briels(2008)]{Dhont}
J.~Dhont and W.~Briels, \emph{Rheol Acta}, 2008, \textbf{47}, 257--281\relax
\mciteBstWouldAddEndPuncttrue
\mciteSetBstMidEndSepPunct{\mcitedefaultmidpunct}
{\mcitedefaultendpunct}{\mcitedefaultseppunct}\relax
\EndOfBibitem
\bibitem[Martens \emph{et~al.}(2012)Martens, Bocquet, and Barrat]{Martens12}
L.~Martens, L.~Bocquet and J.-L. Barrat, \emph{Soft Matter}, 2012, \textbf{8},
  4197\relax
\mciteBstWouldAddEndPuncttrue
\mciteSetBstMidEndSepPunct{\mcitedefaultmidpunct}
{\mcitedefaultendpunct}{\mcitedefaultseppunct}\relax
\EndOfBibitem
\bibitem[Coussot \emph{et~al.}(2002)Coussot, Nguyen, Huynh, and
  Bonn]{Coussot02c}
P.~Coussot, Q.~D. Nguyen, H.~T. Huynh and D.~Bonn, \emph{J. Rheol.}, 2002,
  \textbf{46}, 573\relax
\mciteBstWouldAddEndPuncttrue
\mciteSetBstMidEndSepPunct{\mcitedefaultmidpunct}
{\mcitedefaultendpunct}{\mcitedefaultseppunct}\relax
\EndOfBibitem
\bibitem[Coussot and Ovarlez(2010)]{Coussot10}
P.~Coussot and G.~Ovarlez, \emph{Eur. Phys. Jour. E.}, 2010, \textbf{33},
  183--188\relax
\mciteBstWouldAddEndPuncttrue
\mciteSetBstMidEndSepPunct{\mcitedefaultmidpunct}
{\mcitedefaultendpunct}{\mcitedefaultseppunct}\relax
\EndOfBibitem
\bibitem[Jagla(2007)]{Jagla07}
E.~A. Jagla, \emph{Phys. Rev. E}, 2007, \textbf{76}, 046119\relax
\mciteBstWouldAddEndPuncttrue
\mciteSetBstMidEndSepPunct{\mcitedefaultmidpunct}
{\mcitedefaultendpunct}{\mcitedefaultseppunct}\relax
\EndOfBibitem
\bibitem[Jagla(2010)]{Jagla10}
E.~A. Jagla, \emph{J. Stat. Mech.: Theory Exp.}, 2010,  P12025\relax
\mciteBstWouldAddEndPuncttrue
\mciteSetBstMidEndSepPunct{\mcitedefaultmidpunct}
{\mcitedefaultendpunct}{\mcitedefaultseppunct}\relax
\EndOfBibitem
\bibitem[Dahmen \emph{et~al.}(2009)Dahmen, Ben-Zion, and Uhl]{Dahmen09}
K.~A. Dahmen, Y.~Ben-Zion and J.~T. Uhl, \emph{Phys. Rev. Lett.}, 2009,
  \textbf{102}, 175501\relax
\mciteBstWouldAddEndPuncttrue
\mciteSetBstMidEndSepPunct{\mcitedefaultmidpunct}
{\mcitedefaultendpunct}{\mcitedefaultseppunct}\relax
\EndOfBibitem
\bibitem[Moorcroft \emph{et~al.}(2011)Moorcroft, Cates, and
  Fielding]{Moorcroft11}
R.~L. Moorcroft, M.~E. Cates and S.~M. Fielding, \emph{Phys. Rev. Lett.}, 2011,
  \textbf{106}, 055502\relax
\mciteBstWouldAddEndPuncttrue
\mciteSetBstMidEndSepPunct{\mcitedefaultmidpunct}
{\mcitedefaultendpunct}{\mcitedefaultseppunct}\relax
\EndOfBibitem
\bibitem[Moorcroft and Fielding(2013)]{Moorcroft13}
R.~L. Moorcroft and S.~M. Fielding, \emph{Phys. Rev. Lett.}, 2013,
  \textbf{110}, 086001\relax
\mciteBstWouldAddEndPuncttrue
\mciteSetBstMidEndSepPunct{\mcitedefaultmidpunct}
{\mcitedefaultendpunct}{\mcitedefaultseppunct}\relax
\EndOfBibitem
\bibitem[Chaudhuri \emph{et~al.}(2012)Chaudhuri, Berthier, and
  Bocquet]{Chaudhuri}
P.~Chaudhuri, L.~Berthier and L.~Bocquet, \emph{Phys. Rev. E}, 2012,
  \textbf{85}, 021503\relax
\mciteBstWouldAddEndPuncttrue
\mciteSetBstMidEndSepPunct{\mcitedefaultmidpunct}
{\mcitedefaultendpunct}{\mcitedefaultseppunct}\relax
\EndOfBibitem
\bibitem[Schmitt \emph{et~al.}(1995)Schmitt, Marques, and Lequeux]{Schmitt}
V.~Schmitt, C.~M. Marques and F.~Lequeux, \emph{Phys. Rev. E}, 1995,
  \textbf{52}, 4009--4015\relax
\mciteBstWouldAddEndPuncttrue
\mciteSetBstMidEndSepPunct{\mcitedefaultmidpunct}
{\mcitedefaultendpunct}{\mcitedefaultseppunct}\relax
\EndOfBibitem
\bibitem[Besseling \emph{et~al.}(2010)Besseling, Isa, Ballesta, Petekidis,
  Cates, and Poon]{besseling10}
R.~Besseling, L.~Isa, P.~Ballesta, G.~Petekidis, M.~E. Cates and W.~C.~K. Poon,
  \emph{Phys. Rev. Lett.}, 2010,  105268\relax
\mciteBstWouldAddEndPuncttrue
\mciteSetBstMidEndSepPunct{\mcitedefaultmidpunct}
{\mcitedefaultendpunct}{\mcitedefaultseppunct}\relax
\EndOfBibitem
\bibitem[Fielding(2007)]{Fielding07}
S.~M. Fielding, \emph{Soft Matter}, 2007, \textbf{3}, 1262--1279\relax
\mciteBstWouldAddEndPuncttrue
\mciteSetBstMidEndSepPunct{\mcitedefaultmidpunct}
{\mcitedefaultendpunct}{\mcitedefaultseppunct}\relax
\EndOfBibitem
\bibitem[Nicolas and Barrat(2013)]{Nicolas13b}
A.~Nicolas and J.-L. Barrat, \emph{Faraday Discuss.}, 2013, \textbf{167},
  567--600\relax
\mciteBstWouldAddEndPuncttrue
\mciteSetBstMidEndSepPunct{\mcitedefaultmidpunct}
{\mcitedefaultendpunct}{\mcitedefaultseppunct}\relax
\EndOfBibitem
\bibitem[Chikkadi \emph{et~al.}(2014)Chikkadi, Miedema, Dang, Nienhuis, and
  Schall]{Chikkadi14}
V.~Chikkadi, D.~M. Miedema, M.~T. Dang, B.~Nienhuis and P.~Schall, \emph{Phys.
  Rev. Lett.}, 2014, \textbf{113}, 208301\relax
\mciteBstWouldAddEndPuncttrue
\mciteSetBstMidEndSepPunct{\mcitedefaultmidpunct}
{\mcitedefaultendpunct}{\mcitedefaultseppunct}\relax
\EndOfBibitem
\bibitem[Lemaitre and Caroli(2009)]{Lamaitre09}
A.~Lemaitre and C.~Caroli, \emph{Phys. Rev. E}, 2009, \textbf{103},
  065501\relax
\mciteBstWouldAddEndPuncttrue
\mciteSetBstMidEndSepPunct{\mcitedefaultmidpunct}
{\mcitedefaultendpunct}{\mcitedefaultseppunct}\relax
\EndOfBibitem
\bibitem[Salerno and Robbins(2013)]{SalernoRobbins13}
K.~M. Salerno and M.~O. Robbins, \emph{Phys. Rev. E}, 2013, \textbf{88},
  062206\relax
\mciteBstWouldAddEndPuncttrue
\mciteSetBstMidEndSepPunct{\mcitedefaultmidpunct}
{\mcitedefaultendpunct}{\mcitedefaultseppunct}\relax
\EndOfBibitem
\bibitem[Liu \emph{et~al.}(2015)Liu, E.~E.~Ferrero, and
  Martens]{LiuFerreroarXiv2015}
C.~Liu, J.-L.~B. E.~E.~Ferrero, F.~Puosi and K.~Martens,
  \emph{arXiv:1506.08161}, 2015,  1--6\relax
\mciteBstWouldAddEndPuncttrue
\mciteSetBstMidEndSepPunct{\mcitedefaultmidpunct}
{\mcitedefaultendpunct}{\mcitedefaultseppunct}\relax
\EndOfBibitem
\bibitem[Budrikis and Zapperi(2013)]{Budrikis-PRE2013}
Z.~Budrikis and S.~Zapperi, \emph{Phys. Rev. E}, 2013, \textbf{88},
  062403\relax
\mciteBstWouldAddEndPuncttrue
\mciteSetBstMidEndSepPunct{\mcitedefaultmidpunct}
{\mcitedefaultendpunct}{\mcitedefaultseppunct}\relax
\EndOfBibitem
\bibitem[Lin \emph{et~al.}(2014)Lin, Lerner, Rosso, and Wyart]{Lin-PNAS2014}
J.~Lin, E.~Lerner, A.~Rosso and M.~Wyart, \emph{PNAS}, 2014, \textbf{111},
  14382--14387\relax
\mciteBstWouldAddEndPuncttrue
\mciteSetBstMidEndSepPunct{\mcitedefaultmidpunct}
{\mcitedefaultendpunct}{\mcitedefaultseppunct}\relax
\EndOfBibitem
\bibitem[Puosi \emph{et~al.}(2015)Puosi, Oliver, and
  Martens]{Puosi-SoftMatter2015}
F.~Puosi, J.~Oliver and K.~Martens, \emph{arXiv:1501.04574}, 2015,  1--8\relax
\mciteBstWouldAddEndPuncttrue
\mciteSetBstMidEndSepPunct{\mcitedefaultmidpunct}
{\mcitedefaultendpunct}{\mcitedefaultseppunct}\relax
\EndOfBibitem
\bibitem[Bouchbinder and Lo(2008)]{STZbouchbinder}
E.~Bouchbinder and T.-S. Lo, \emph{Phys. Rev. E}, 2008, \textbf{78},
  026119\relax
\mciteBstWouldAddEndPuncttrue
\mciteSetBstMidEndSepPunct{\mcitedefaultmidpunct}
{\mcitedefaultendpunct}{\mcitedefaultseppunct}\relax
\EndOfBibitem
\bibitem[Hebraud and Lequeux(1998)]{Hebraud}
P.~Hebraud and F.~Lequeux, \emph{Phys. Rev. Lett.}, 1998, \textbf{81},
  2934\relax
\mciteBstWouldAddEndPuncttrue
\mciteSetBstMidEndSepPunct{\mcitedefaultmidpunct}
{\mcitedefaultendpunct}{\mcitedefaultseppunct}\relax
\EndOfBibitem
\bibitem[Risken(1996)]{Risken}
H.~Risken, \emph{The Fokker-Planck Equation: Methods of Solution and
  Applications}, Springer, 1996\relax
\mciteBstWouldAddEndPuncttrue
\mciteSetBstMidEndSepPunct{\mcitedefaultmidpunct}
{\mcitedefaultendpunct}{\mcitedefaultseppunct}\relax
\EndOfBibitem
\end{mcitethebibliography}
\bibliographystyle{rsc} 
\end{document}